\documentclass[conference]{IEEEtran}
\IEEEoverridecommandlockouts

\usepackage{cite}
\usepackage{amsmath,amssymb,amsfonts}
\usepackage{algorithmic}
\usepackage{graphicx}
\usepackage{textcomp}
\usepackage{xcolor}
\def\BibTeX{{\rm B\kern-.05em{\sc i\kern-.025em b}\kern-.08em
    T\kern-.1667em\lower.7ex\hbox{E}\kern-.125emX}}

\usepackage{multirow}
\usepackage{xspace}
\usepackage{todonotes}

\usepackage{url}

\newcommand\proto{\textit{H-Quorum}\xspace}

\makeatletter 
\newcommand{\linebreakand}{%
  \end{@IEEEauthorhalign}
  \hfill\mbox{}\par
  \mbox{}\hfill\begin{@IEEEauthorhalign}
}
\makeatother 

\begin{document}

\title{Resilient and Secure Programmable System-on-Chip Accelerator Offload}

\author{\IEEEauthorblockN{In\^{e}s Pinto Gouveia\IEEEauthorrefmark{1},
Ahmad T. Sheikh\IEEEauthorrefmark{2}, Ali Shoker\IEEEauthorrefmark{3},\linebreakand
Suhaib A. Fahmy\IEEEauthorrefmark{4} and Paulo Esteves-Verissimo\IEEEauthorrefmark{5}}
\IEEEauthorblockA{Computer, Electrical and Mathematical Sciences and Engineering Division (CEMSE),\linebreakand
King Abdullah University of Science and Technology (KAUST)\\
Thuwal 23955-6900, Kingdom of Saudi Arabia\\
Email: \IEEEauthorrefmark{1}ines.pintogouveia@kaust.edu.sa,
\IEEEauthorrefmark{2}ahmad.sheikh@kaust.edu.sa,
\IEEEauthorrefmark{3}ali.shoker@kaust.edu.sa,\linebreakand
\IEEEauthorrefmark{4}suhaib.fahmy@kaust.edu.sa
\IEEEauthorrefmark{5}paulo.verissimo@kaust.edu.sa}}

\maketitle

\begin{abstract}
Computational offload to hardware accelerators is gaining traction due to increasing computational demands and efficiency challenges. Programmable hardware, like FPGAs, offers a promising platform in rapidly evolving application areas, with the benefits of hardware acceleration and software programmability. Unfortunately, such systems composed of multiple hardware components must consider integrity in the case of malicious components. In this work, we propose Samsara, the first secure and resilient platform that derives, from Byzantine Fault Tolerant (BFT), protocols to enhance the computing resilience of programmable hardware. Samsara uses a novel lightweight hardware-based BFT protocol for Systems-on-Chip, called H-Quorum, that implements the theoretical-minimum latency between applications and replicated compute nodes. To withstand malicious behaviors, Samsara supports hardware rejuvenation, which is used to replace, relocate, or diversify faulty compute nodes. Samsara's architecture ensures the security of the entire workflow while keeping the latency overhead, of both computation and rejuvenation, close to the non-replicated counterpart.
\end{abstract}


\section{Introduction}
\label{sec:introduction}

The trend of offloading computing to hardware accelerators is gaining more traction in emerging systems and applications such as cyber-physical systems (CPS), Internet-of-Things (IoT) services, automation and space applications~\cite{Safarulla2014, Shashidhara2020, Feng2022}. The purpose is often both seeking accelerated computing and enhanced security. Hardware acceleration is a result of building application-specific logic, e.g., matrix multiplication and cryptographic operations, precluding OS or software interruptions or locking, and taking advantage of fine grained parallelism; whereas security leverages the hardware immutability properties to provide guarantees like tamper-resistant data processing, isolation and building security abstractions~\cite{SGX2016, ARMTrustZone, kinney2006trusted, sabt2015trusted, Distler2016, hw-assisted-2018scalable,MinBFT2013}. To continuously support new use-cases while reducing the burden of hardware fabrication of a monolithic system (costly and slow), these systems are often (1) assembled and deployed as \textit{Multi-Processor System-on-Chip} (MPSoCs) that makes use of modular cheap \textit{Commercial-of-the-shelf} (COTS) components~\cite{ZynqMPSoC,shoker2023path}; and (2) leverage the hardware programmablity features of general-purpose reconfigurable hardware like FPGAs and CGRAs~\cite{karandikar2018firesim, esp2020, singh2000morphosys}. Unfortunately, this raises new computing integrity challenges against intrusions and faults.

Reconfigurable hardware fabrics allow new computing functionality to be loaded after fabrication~\cite{boutros2021fpga}, by means of a binary file (usually called the \textit{bitstream}) that maps an architecture or module (we call a \emph{Tile} henceforth) onto the fabric. While this gains the intrinsic performance and security properties of hardware after loading, it inherits some of the mutability weaknesses of software prior to loading (given the nature of the binary file), which can be prone to intrusion and benign faults~\cite{mal2014hardware,duncan2019fpga,ender2020unpatchable,ender2022cautionary,Feng2022}. We noticed that little has been done to ensure computing integrity when a tile is faulty or malicious (\textit{Byzantine}~\cite{BFT1982}). Indeed, literature has addressed this issue through several complementary approaches, e.g., focusing on encryption, isolation, and memory obfuscation~\cite{BitStreamEncryption,zhang2014secure,avramenko2019rtos}; however, these approaches assume tiles are trusted or rely on software and, thus, mutable implementations. For instance, encrypting the bitstream~\cite{BitStreamEncryption} prevents tampering with it at rest, e.g., in memory; however, it neither withstands a contaminated bitstream at the development phase (which is can occur due to the lengthy hardware development process and the use of many tools), nor \textit{Network-on-Chip} (NoC) attacks under execution~\cite{charles2021survey}. On the other hand, containment and memory randomization~\cite{zhang2014secure} ensure the execution of deployed co-existing
tiles do not interfere, but do not give any guarantees on the output integrity if tiles are Byzantine, e.g., corresponding to a vulnerable bitstream. Finally, \textit{Triple-Modular-Replication} (TMR) is often used as a voting mechanism to ensure computing integrity~\cite{lyons1962use}, but it is not adequate for malicious adversaries. 

In this paper, we introduce Samsara, the first resilient and secure computing platform architecture for programmable MPSoCs. Samsara hinges on hardware reconfigurability to ensure computing integrity by employing state-machine replication (SMR)~\cite{fischer1985impossibility} and rejuvenation of tiles, managed through a novel lightweight hardware-based Byzantine agreement protocol, called \proto{}. \proto{} is optimized for simplicity and low latency to make it feasible for hardware offload. When \proto{} detects a fault or delay, it recovers via a rejuvenation process that is seamless to the application. Rejuvenation is used to reload (from a bitstream library) a tile facing transient faults or glitches, replace a malicious or faulty tile with a diversified version~\cite{makni2016comparison} to improve independence of failures, or relocate the tile to another location to avoid underlying glitches or compromised network routes. Diversity of bitstreams is possible by means of different logic module configurations at design time, through design with hardware description languages (HDLs) or open-source architectures and frameworks like those of RISC-V~\cite{waterman2014risc,waterman2014risc,marques2021lock}. Samsara's architecture is designed to maintain the security of the entire workflow including storing encrypted bitstreams, booting, execution, and rejuvenation.

One of the main challenges addressed by Samsara is circumventing the known complexity and delay of Byzantine agreement, which would represent a significant overhead for hardware accelerators. Hence, we optimize H-Quorum for simplicity and latency by inspiring from Quorum~\cite{guerraoui2010next}, which is known to be the simplest classical Byzantine agreement protocol with the lowest theoretical latency possible~\cite{guerraoui2010next, adapt:2015}. Nevertheless, Quorum is known to be impractical in Internet-based settings, as it assumes that clients are not Byzantine and it recovers into a heavy-weight backup phase under faults (see details in Section~\ref{sec:arch}). We circumvent these limitations by making smart choices: Samsara employs an \textit{Application Specific Integrated Circuit} (ASIC) \textit{Controller} that plays the role of a trusted "semi-leader" replica: it acts as a leader by mediating the application requests sent to the replicated tile and collects votes, while not participating in the computation phase. This is key to ensure that the Controller exhibits a small fingerprint (a few hundred LoC --- of a hardware description language --- in our case) to be easily verified, and, thus, considering it a trusted-trustworthy hardware component is a valid assumption.

Another main challenge in Samsara is to ensure secure rejuvenation. Rejuvenation is done while the system is running by reloading a new bitstream to the reconfigurable fabric via booting scripts. To make sure intruders cannot modify, initiate, or interrupt rejuvenation, our MPSoC architecture makes use of a simple microprocessor in which only the Controller can run these scripts. This makes sure that \proto{} maintains the state integrity across rejuvenation, including a fast shared-memory state-transfer mechanism, detailed in Section~\ref{sec:phases}. We provide a systematic proof sketch in Appendix~\ref{sec:proofs} to ensure the correctness of the entire workflow phases.

As a proof of concept, we implemented Samsara on a Xilinx ZCU102 FPGA SoC~\cite{zcu102}. Our evaluation of accelerator applications shows that Samsara's latency is slightly higher than a non-replicated accelerator, and up to 35.9\% lower than state-of-the-art hardware-based Byzantine agreement counterparts, like \emph{iBFT}~\cite{pinto2022architectural} and a shared-memory implementation of \emph{MinBFT}~\cite{MinBFT2013}. Additionally, the evaluation shows that the rejuvenation time is negligible, and  99.89\% faster than rebooting the whole platform.

The rest of the paper is organized as follows: Section~\ref{sec:sys-threat-model} discusses the Samsara's system and threat models, as well as background on MPSoCs and reconfiguration. Samsara's architecture and \proto{} are presented in Section~\ref{sec:proposed-framework}. We then provide the evaluation of our proof of concept on an ZCU102 FPGA in Section~\ref{sec:evaluation}, and discuss related work in Section~\ref{sec:related-work}. Finally, we conclude in Section~\ref{sec:conclusion}. We provide correctness and liveness proof sketches in Appendix~\ref{sec:proofs}, for the reader's convenience.

\section{System and Threat Models}
\label{sec:sys-threat-model}

\subsection{System Model}
\subsubsection{Generic Programmable MPSoC}
We consider a generic programmable \textit{Multi-Processor System on Chip} (MPSoC) composed of: (1) a processor portion, called the \textit{Processing System} (PS) that has one or more processing cores, used to run application software, alongside the basic I/O and peripherals; (2) a reconfigurable portion (e.g., FPGA, CGRA), used to deploy offloaded functionality, i.e., accelerated tasks. Examples of commercial devices tht fit this model include the AMD/Xilinx Zynq-7000~\cite{zynq7000} and Zynq UltraScale+~\cite{ZynqMPSoC} and Intel/Altera Agilex 7 SoC. All portions of the MPSoC are connected via an underlying reliable hardware Network-on-Chip (NoC) or network bus, such as AMBA AXI or PCIe~\footnote{The main difference between a NoC and a bus pertains to the fact that the NoC uses a conceptually point-to-point approach, while a bus tends to be multipoint.}. The choice of communication medium depends on the platform's fine-grained architecture and application requirements (e.g., bandwidth). Channels may slightly delay circulated messages or data but will eventually deliver them as long as they are not under attack. 

\subsubsection{Samsara COTS}
In the context of Samsara, we propose an additional \textit{application-specific integrated circuit} (ASIC) component, as part of the MPSoC PS, which implements the \textit{Controller} of \textit{Samsara} (as explained in Section~\ref{sec:arch}); and a simple COTS microprocessor, \textit{MP-Boot}, to handle booting and configuration of the reconfigurable portion, (see Section~\ref{sec:arch})). We propose that the MPSoC is supplied with volatile memory and hardware-assisted \textit{Tamper-Resistant Storage} (TRS) (e.g., \textit{TPM}~\cite{tpm} or \textit{Hardware Credential Storage}~\cite{sharma2007onboard}), used for tamper-proof storage and compute. Modern MPSoCs allow addition of specific COTS, modules, or \textit{microchips} as we do~\cite{10.1145/3503222.3507742}. Future MPSoC designs leveraging chiplets also provide this flexibility.

In addition to this, as detailed in Section~\ref{sec:arch}, the reconfigurable portion is divided into partitions, i.e., \emph{tiles}~\cite{waingold1997baring}, that encapsulate compute functions. A tile is an abstraction of the logic modules containing offloaded and/or accelerated tasks, placed on dynamically reconfigurable hardware regions. Samsara uses these tiles as \proto{} replicas.

\subsubsection{Reconfiguration Implementation}
There are several possible implementations for the reconfigurable portion. For instance, \textit{FPGA Programmable Logic} (PL) can be used to instantiate hardware accelerators in two phases: (1) loading \textit{full bitstreams} that define the base infrastructure, e.g., determining the number of tiles locations, networking (called routing), and other components; and (2) \textit{partial bitstreams} for reconfiguration at runtime, used to deploy new compute logic. Alternatively, CGRAs~\cite{liu2019survey} consist of a large array of function units (FUs) interconnected by a mesh style network; they offer more coarser-grained configuration than FPGAs. Yet another possibility is to use multiple connected GPUs or other diverse FUs, e.g., via a secure implementation of PCIe~\cite{sec_pcie}.

For the remainder of this paper, we shall focus on an FPGA-based implementation of the reconfigurable portion. For this, we first present a brief overview of MPSoC and FPGAs to ease the understanding of the rest of the remaining sections.

\subsection{Background on MPSoC and FPGA}
\label{sec:background}

This section provides a brief background on programmable Multi-Processor Systems-on-Chip (MPSoC) and FPGA Partial Reconfiguration (PR).

\subsubsection{MPSoC Architecture}
\label{sec:zynq-mpsoc}
An MPSoC is a System-on-a-Chip (SoC) which includes multiple processors, often used in embedded devices. A typical modern MPSoC~\cite{ZynqMPSoC} includes multiple fabricated processing cores (called \textit{hard} cores) in a Processing Side (PS) and programmable hardware, based on Field Programmable Gate Array (FPGA) technology. The latter can be reconfigured with arbitrary custom-hardware logic after fabrication.

The cost of FPGA development and deployment is negligible for low to medium volumes. However, per-device cost is expensive. Application Specific Integrated Circuits (ASICs) are chips with immutable logic circuits. They are preferred for large scale manufacturing due to cheaper per-device cost after amortizing the high non-recurring engineering (NRE) costs.

\subsubsection{FPGAs}
FPGAs possess reconfigurable programmable logic (PL), consisting of fabric that can be programmed according to a design that is mapped into a configuration file called a \emph{bitstream}. A common case is to design accelerators or softcores, and connect them to other modules such as memory controllers and I/O. The PL can be configured directly through an external interface or via the MPSoC's PS. 
The PL consists of a sea of building blocks such as Configurable Logic Blocks (CLB), programmable routing, I/O banks, Block RAMs (BRAMs), etc. Designed modules, e.g., accelerators or softcores, and their communication medium are mapped into the aforementioned components.

Traditionally, when the PL is programmed, it becomes immutable, i.e., the free and configured regions of the PL cannot be further changed without reconfiguring the whole fabric. However, dynamic partial reconfiguration (PR) allows partitions of the PL to be reconfigured at runtime~\cite{vipin2018fpga}, without mutating the rest of the design. Thus, a full bitstream configures the whole PL, while partial bitstreams modify only the specified partition(s) without compromising the integrity of applications running on the remaining portions of the PL. One of the biggest advantages of PR is the ability to time multiplex the underlying silicon for various tasks.

To design a system using PR is it necessary to \textit{floorplan} the design: the PL fabric is spatially divided into \textit{Reconfigurable Partitions (RPs)} that are to contain dynamically \textit{Reconfigurable Modules} (RMs). When using PR, floorplans contain a static partition, containing the design modules that cannot or should not be dynamically reconfigured~\footnote{These can, however, be reconfigured by reconfiguring the whole PL.}; and one or more RPs. The number and placement of these RPs cannot be changed at runtime, meaning that, in order to change the number of RPs used and the type of RMs they can hold, a full bitstream (configured with a new floorplan) must be loaded to the PL. Compatible RMs, e.g., diverse accelerators, can, however, be swapped at runtime using only partial bitstreams.

We now present the threat model we assume, based on MPSoC and FPGAs.

\subsection{Threat Model}
\label{sec:threat}

In this section we address Samsara's hardware, software and network/bus threat models (see Samsara's architecture in Fig.~\ref{fig:samsara} for clarity). The threat model focuses on compute integrity and availability. Confidentially is out of scope for this paper.

\subsubsection{Samsara Controller and MP-Boot}
Samsara's \emph{Controller} is a simple and easily-verifiable trusted-trustworthy ASIC and is, therefore, assumed to be tamper proof. Similarly, the microprocessor \emph{MP-Boot} has a small footprint and is used only to run booting and reconfiguration software upon a signal from the \emph{Controller}. This software is stored in encrypted form in, e.g., the TPM. Therefore, this software is assumed to preserve its integrity and not be called by another entity. Both the \emph{Controller} and \emph{MP-Boot} are detailed in Section~\ref{sec:proposed-framework}.

\subsubsection{PL Hardware Logic}
We assume a threat model where an adversary can inject hardware trojans~\cite{chakraborty2013hardware,swierczynski2017bitstream} or backdoors (e.g., in coalition with PS software~\cite{ancajas2014fort}) in PL Tiles, during the design cycle~\cite{duncan2019fpga, mal2014hardware}. That is, we assume tiles in the PL and, thus, their bitstreams may contain malicious functionality that impacts \emph{integrity}. Additionally, we assume the possibility of non-malicious threats caused by unintended vulnerabilities created at design time~\cite{duncan2019fpga}. We do not focus on attacks that have been successful in breaking the \emph{confidentiality} of bitstreams~\cite{ender2020unpatchable,ender2022cautionary} while they are at rest in memory.

To boost resilience against a compromised tile, several redundant tiles are run simultaneously in the PL following the \proto{} agreement protocol --- explained in Section~\ref{sec:phases}. We assume that a majority of these tiles are not simultaneously compromised in a short time window, which means they have diverse configurations or implementations to avoid common-mode failures. For example, diverse configurations of hardware modules are a possibility that is currently supported by some vendor tools, while different implementations are also possible via hardware description languages (HDLs) or with the support of the RISC-V instruction set architecture and respective frameworks~\cite{waterman2014risc,marques2021lock}. We also assume that tiles have a level of containment or isolation, by having the base PL design use methods like Xilinx's Isolation Design Flow (IDF)~\cite{xilinx_idf}, which provides fault containment at the FPGA module level, so that an anomaly or vulnerability may not affect other modules of the PL directly.

\subsubsection{General Hardware}
\emph{Application Cores} may fail arbitrarily since these do not have access to the \emph{Controller}, \emph{MP-Boot}, the TPM, or the PL. For further isolation guarantees, solutions like~\cite{gouveia2022behind} can be used for flexible access control. We do not assume any side-channel attacks on the platform, the failure of the FPGA fabric itself, nor MPSoC-wide failures. This assumption is reasonable since FPGAs and MPSoCs in general are high-end hardware that are subject to rigorous testing. Still, resilient clocks~\cite{schmid2010decentralised} mitigate some chip-wide common-mode faults and the recent trend towards interconnected chiplets further improves the physical decoupling of MPSoC components.

\subsubsection{Software}
Similarly, the attacker can compromise any application software that would be executed in the PS's Application Cores (see Section~\ref{sec:arch}) or in PL soft cores (if used as tiles). However, as we shall explain later, malfunctioning PS application software does not interfere with Samsara, and software running on the PL (if used) has no means to cause trouble beyond providing incorrect output, which Samsara addresses with \proto{}. Furthermore, to safeguard Application Cores further, their core-to-NoC interface can be augmented with solutions such as~\cite{gouveia2022behind}.

\subsubsection{Network-on-Chip}
Moreover, via a compromised tile, the adversary may compromise the NoC by dropping or modifying exchanged messages~\cite{sepulveda2017towards,srivastava2005performance}, namely by means of software-hardware coalition~\cite{ancajas2014fort}. The adversary cannot, however, spoof, pretending to be another tile, as shown in Annex~\ref{sec:proofs}.

\section{Samsara Resilient Computing Platform}
\label{sec:proposed-framework}

\begin{figure}[t]
\centering
\includegraphics[width=0.8\columnwidth]{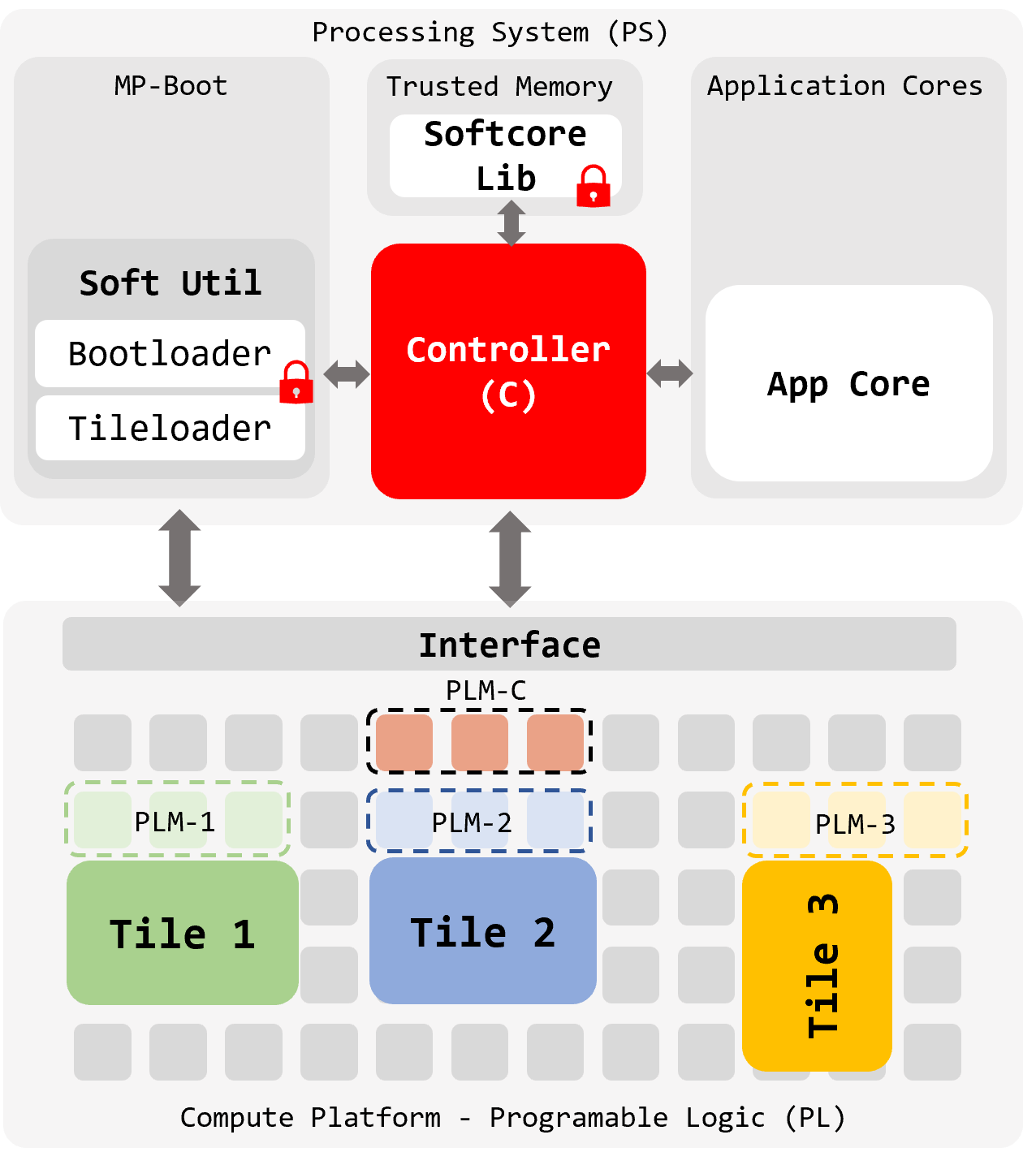}
\caption{The architecture of \textit{Samsara} Rejuvenation Fault and Intrusion Tolerant Framework.
The three tiles are spawned from three known softcores~\cite{makni2016comparison} to demonstrate diversity.}
\label{fig:samsara}
\end{figure}

\subsection{Architecture}
\label{sec:arch}
We present a high-level architecture of \emph{Samsara} in Figure~\ref{fig:samsara}. Samsara's architecture is composed of three main components: the \emph{Controller}, the \emph{Compute Platform}, and \emph{MP-Boot Utilities}.
The \emph{Controller} manages the \emph{Compute Platform} in the PL and interfaces the application requests from the \emph{Application Cores} with it. Having such a critical role, the \emph{Controller} is implemented in hardware (ASIC) that cannot be tampered with. To enable different configurations and extensions, the Controller stores default PL configurations and security keys in Tamper-Resistant Storage (TRS). The \emph{Compute Platform} implements the compute \emph{Tiles}, i.e., accelerators and critical functions, to be used by the applications. It is composed of FPGA-based tiles in the PL, loaded from bitstreams, encrypted and stored in the Softcore library in the PS's memory and later authenticated. Encryption provides basic design security to protect the design from copying or reverse engineering (which is not the focus of this paper), while authentication provides assurance that the bitstream provided for configuration is the unmodified bitstream created by an authorized user. Authentication verifies both data integrity and authenticity of the bitstream. Authorized users can still have corrupt or malicious bitstreams and these can also be subject to faults, hence the need for PL-side Rejuvenation, as outlined in Section~\ref{sec:phases}.

With the benefit of reconfigurable hardware, tiles are subject to updates over time, e.g., to modify functionality, which makes them less secure than fixed function ASIC tiles. Consequently, the Compute Platform supports active replication of tiles to mask and detect faulty or misbehaving ones. This is protected and managed by the Controller. In addition, \emph{Samsara} makes use of \textit{Software Utilities}, i.e., the \emph{Bootloader} and \emph{Tileloader} (see Fig.~\ref{fig:samsara}), that are used occasionally by the \emph{Controller} to manage and rejuvenate faulty tiles. These utilities are software-based components that should only interact with the PL at booting and reconfiguration time (detailed in the next section), respectively. Therefore, they are retained in memory in an encrypted form and are executed in a dedicated processor (\emph{MP-Boot}) that is the only software processing unit with PL access. Furthermore, \emph{MP-Boot} is not available for user-level application usage, to prevent malicious code from invoking the API that loads PL bitstreams.

\begin{figure}[t]
\centering
\includegraphics[width=.8\columnwidth]{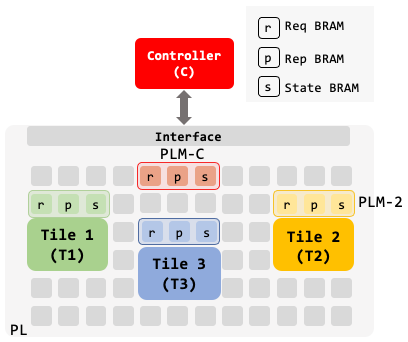}
\caption{Memory access in Samsara: each entity has \textit{R/W }access to its own PL Memory (PLM). C has \textit{R} access to all PLMs. Each Tile has \textit{R} access to PLM-C.}
\label{fig:memory}
\end{figure}

In our architecture, the communication between the Controller and the Compute Platform Tiles can be done, e.g., 1) through a bus such as PCIe, or 2) through shared PL memory - BRAMs. While 1) would require the use of signatures for authentication, 2) does not, as we explain in Annex~\ref{sec:proofs}. We shall use 2) for the remainder of the paper. As seen in Fig.~\ref{fig:memory}, the Controller holds a BRAM in the PL to which it has read/write access. Here, it writes the Requests and keeps the Log of executed Requests for later state transfer after rejuvenating a tiles (in the case of stateful applications). Tiles have read-only access to the Controller's BRAM in order to prevent malicious ones from modifying its contents. Then, tiles hold a BRAM of their own, to which they have read/write access and the Controller has read-only access. In this BRAM, tiles place their Replies as well as their local Log. Requests and Replies are written up to a maximum, after which a checkpoint is taken and the BRAMs reset to give space to new rounds of Requests/Replies. Checkpoints are saved in an on-chip SRAM outside the PL and can be fetched if needed by the Controller. Reset of the BRAMs can be triggered by the Controller or by a simple Reset IP in the PL.

The Controller's memory can instead be kept outside the PL as regular SRAM or as Tamper-Resistant Storage (TRS), which could improve security due to being outside reconfigurable logic, but, depending on the specific implementation (bus, type of storage), could as well increase access times.

\subsection{Operational Phases}
\label{sec:phases}

\subsubsection{Overview}

Samsara operates in three phases: \textit{Bootstrapping}, \textit{Execution}, and \textit{Rejuvenation}. In a nutshell, in the Bootstrapping phase, the Controller launches the Compute Platform. It executes the \textit{Bootloader} utility that loads the main PL bitstream (i.e., the floorplan of the FPGA) and then the \textit{Tileloader} utility following the configuration stored in the TRS. The configuration decides the types/versions of tiles to load from the Softcore Library as partial bitstreams, how many tile replicas are run, when and how to rejuvenate tiles, etc. When the Compute Platform is ready for use by the applications, the \textit{Execution} phase begins. In this phase, the Controller receives requests from the applications, assigns each request a unique ID, and sends them to all the tiles of the Compute Platform, by running the lightweight quorum-based Byzantine agreement \proto{} (detailed next). The tiles execute the same request simultaneously\footnote{Simultaneous here does not mean in lock step.} and reply to the Controller. The latter verifies if a majority of the received responses are matching and forwards the reply back to the application. In case there is any mismatch or delay detected, the corresponding tile is to be replaced. This announces the launch of the \textit{Rejuvenation} phase. In this phase, the Controller destroys the faulty tile and replaces it with another one of the same type (i.e., refreshing it) or by another diverse version. This follows a policy defined by the Controller. Rejuvenation ends by completing the state transfer to the newly loaded tile if the application is stateful.

Next, we explain the three phases of Samsara in detail.

\begin{table}[t]
\footnotesize
    \centering
            \caption{Controller Configuration.}

    \begin{tabular}{p{1.2cm}p{6.7cm}}\hline
      \textbf{Attribute} & \textbf{Description}\\
    \hline
    \textit{softcore} & compute softcore types (stored as bitstreams) \\
    \textit{version} & softcore diverse versions that are functionally similar but exhibit different design, implementation, or configuration. \\
    \textit{min-tiles} & minimum number of tiles allowed (could be defined by the agreement protocol used).\\
    \textit{max-tiles} & maximum number of tiles allowed, limited by PL capacity and by floorplanning\footnotemark choices at design time.\\
    \textit{stateful} & boolean is \textit{True} if the application is stateful.\\ 
     \textit{rejuv-policy} & defines the rejuvenation modes: Refresh/Diversify, Replace/Relocate, Scale-in/Scale-out, or Reactive/Proactive.\\\hline
    \end{tabular}
    \label{tab:config}
\end{table}

\footnotetext{Floorplanning refers to the set of physical constraints used to control how logic is placed in the PL. The definition of dynamically reconfigurable zones is done at this stage and defines the containers these partitions can be in.}
    
\subsubsection{Bootstrapping Phase} This phase is only launched when the MPSoC is started. It aims at preparing the Compute Platform through loading the necessary main and partial (i.e., Tile) bitstreams to the PL section with the following steps:
\begin{enumerate}
    \item Controller sets its status: $\mathbf{status}\leftarrow Loading$ and accepts no requests from applications.
    \item Controller verifies and launches the \textit{Bootloader} in the dedicated microprocessor. This loads the main bitstreams to the PL. The main bitstream represents the basic configuration on which Tile bitstreams are loaded afterwards into specific containers defined by the main one.
    \item Controller verifies and launches the \textit{Tileloader} using the stored configuration in the TRS. The Controller passes the configuration parameters $\mathbf{Config}$ to the Tileloader. The retained configurations in the TRS are described in Table~\ref{tab:config}.
    \item Tileloader loads the bitstreams to the PL and notifies Controller when the Tiles are ready. This avoids the case when the latter is unresponsive or faulty. The Controller sets a timer while waiting for the Tiles \textit{status} to become \textit{Ready}.
    \item Tiles notify the Controller when \textit{Ready}. This asserts the info sent by the Tileloader in case it is unresponsive or faulty. 
    \item If the Controller received the expected number of Ready messages from the tiles, as defined in $\mathbf{Config}$ before the timer's expiry, it sets its $\mathbf{status}\leftarrow Ready$ and starts accepting application requests. Otherwise, the Controller launches the \textit{Rejuvenation} phase in \textit{partial-mode} if a minority of Tiles are faulty or slow, or in full-mode otherwise.
\end{enumerate}

\subsubsection{Execution Phase} 
\label{sec:exec_phase}
\begin{figure}[t]
\centering
\includegraphics[width=0.7\columnwidth]{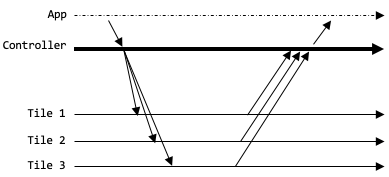}
\caption{Message exchange in \textit{Samsara}'s \proto \textit{Execution} phase.}
\label{fig:msg-pattern}
\end{figure}

This phase is dedicated to the execution of application requests through a novel lightweight quorum-based \textit{Byzantine} agreement protocol, we call it \proto. \proto has a simple message exchange pattern, depicted in Fig.~\ref{fig:msg-pattern}, that is tailored for low latency. In a nutshell, an application sends its requests to the Controller that mediates the requests and responses with the compute Tiles. The Controller sends each request directly to all Tiles and collects their responses. A reply is sent to the application if a majority, i.e., $f+1$ out of $2f+1$, of responses match. Otherwise it recovers by launching the rejuvenation phase. 

\proto is inspired by \textit{Quorum}~\cite{guerraoui2010next}, that is proposed for Internet-based distributed services settings. \textit{Quorum} follows a single round-trip direct messaging pattern between a client and all the $3f+1$ replicas, assuming $f$ faulty replicas, thus making it the Byzantine agreement protocol with the lowest theoretical latency~\cite{guerraoui2010next,adapt:2015}. This makes it the preferred choice in our case since low latency is paramount in hardware accelerator applications. 
Nevertheless, Quorum has several shortcomings that make it unfeasible for Internet-based settings, and actually impede its adoption~\cite{guerraoui2010next,adapt:2015}. In particular, (i) Quorum assumes the client is trusted since it directly sends requests to all replicas and collects their responses, i.e., without a primary replica as in classical protocols~\cite{PBFT1999,zyzzyva2010}; (ii) requiring $3f+1$ replicas to maintain safety and liveness under partially-synchronous networks, which is deemed costly; and (iii) requiring a recovery phase to a backup protocol (e.g., PBFT~\cite{PBFT1999}) with an expensive state transfer mechanism due to logs' matching, necessary for safe view-change under failures. Fortunately, by leveraging the hardware environment, \proto tweaks Quorum in order to bypass these shortcomings as follows.\\

\noindent \textbf {A. Controller acts as a trusted half-primary.}
\proto avoids the trusted client issue by having the Controller mediating the messages between the application and the Tiles. Being an ASIC hardware, the Controller is arguably highly trusted provided that its footprint is not big (as we convey in Section~\ref{sec:evaluation}) in order to be easily verifiable. For this, we decided to simplify the Controller that can be observed as "half-primary": it mediates the messages and performs the log matching, but it does not compute the requests as the Tiles do (since computation logic can have a large footprint in accelerators). \\

\noindent \textbf {B. Requiring a majority of correct tiles.}
\proto requires only $2f+1$ tiles to tolerate a maximum of $f$ faulty tiles to maintain liveness and safety. This is possible since the communication in Samsara is hardware-based, which provides some network containment and thus exhibiting some network synchrony. This allows us to make some time bounds on responses before launching the rejuvenation phase, contrary to Quorum that cannot differentiate between a Byzantine replica and a slow or faulty network. Even in the case where a faulty Tile is slow or attempts to launch a DoS attack~\cite{charles2021survey} on the network bus or jam the Controller, the latter can detect this easily and launch the rejuvenation phase that can change the faulty Tile and/or the network routes (see next). \\

\noindent \textbf {C. Safe state-transfer upon recovery.}
\proto's state-transfer is simple as it does not require log-matching, contrary to Quorum. Indeed, Samsara's Controller has a restricted memory in the PL, i.e, PLM-C in Fig.~\ref{fig:samsara}, in which it retains the state and logs, that are equivalent to those of \textit{correct} Tiles.  This makes state-transfer fast and importantly preserves the correct state across rejuvenation phases (which is analogous to \textit{view-change} in Quorum).\\

We now present the steps of \proto as follows:
\begin{enumerate}
    \item An application $A$ sends a request $req$ to the Controller $C$.
    \item $C$ creates a message $m=uid, req, H(req)$, where $uid$ is the message \textit{unique ID} representing an incremented sequence number safely assigned by the Controller, and $H(Req)$ is the hash digest of $Req$. $C$ sends $m$ to each tile $T_i$ by placing $m$ message in the read-shared PLM-C memory. The Controller sets a timer $timer_i$ and waits for a reply from at least $f+1$ tiles before expiry.
    \item Upon the receipt of $m$, a tile $T_i$ accepts $m$ after verifying its hash digest and $uid$. Then, $T_i$ computes $req$ and puts the output $rep$ in a response message $r=uid, rep, H(rep), tid$ that uses the corresponding request $uid$, as well as $T_i$'s ID $tid$. The latter "sends" $r$ to $C$ by placing it in its shared $PLM_i$ slot to be read by $C$.
    \item For each tile $T_i$, $C$ reads the reply $r_i$ from the corresponding $PLM_i$, it verifies its hash digest and $uid$. Then, $C$ tries to check if a majority of these replies are matching as follows:
    \begin{enumerate}
        \item If $C$ received matching (i.e., having equivalent output) replies $rep$ from all the $2f+1$ tiles before the $timer_c$ expires, it forwards the $rep$ to the application $A$. In addition, if $A$ is a stateful application, $C$ \textit{temporally} retains the $req|rep$ in its allocated PL $PLM_C$ slot to be used for state transfer under faults.
        \item If $C$ received only $f+1$ matching replies $rep$ before $timer_c$ expires, it forwards $rep$ to $A$; however, it also launches the \textit{Rejuvenation} phase with \textit{partial-mode} to replace the faulty or slow tile (see next). In addition, if $A$ is stateful, $C$ \textit{temporally} retains the $req|rep$ in its allocated PL $PLM_C$ slot to be used for state transfer under faults.
        \item Otherwise, if less than $f+1$ matching responses $rep$ are received before $timer_c$ expires, $c$ launches the \textit{Rejuvenation} phase with \textit{full-mode} to replace all the tiles and their (network) routes (see next).
        \item Finally, if agreement was successful, $C$ marks the request as done.
    \end{enumerate}
   
\end{enumerate}

\subsubsection{Rejuvenation Phase} This phase is launched by running the \textit{Tileloader} to carry on a \textit{partial-mode} or \textit{full-mode} replacement of the Compute Platform. \textit{Partial-mode} happens through reloading new softcores to the PL, e.g., to refresh or diversify tiles; whereas the \textit{full-mode} rejuvenates the tiles as well as their routes (network). Both stateless or stateful applications can be considered. For the former, no state transfer upon rejuvenation is required, while for the latter rejuvenated replicas read or copy the state saved in PLM-C, depending on whether they need a local copy. Examples of stateful applications include, e.g., image processing. Rejuvenation steps and modes are explained as follows: 
\begin{enumerate}
    \item The Controller $C$ changes its status to $\mathbf{status}\leftarrow Loading$ and stops accepting application requests. 
    \item $C$ invokes the \textit{Tileloader} in two possible modes: (i) \textit{Partial-mode} where $C$ launches the \textit{Tileloader} to flush the faulty or slow tiles and their corresponding BRAMs in the PLMs. It then reloads the same or diverse softcores as new tiles (see next). In the (ii) \textit{full-mode}, $C$ invokes $Tileloader$ on the entire Compute Platform to be rejuvenated, meaning that all PL resources (softcores, BRAMs, Routes, Registers, etc.) are destroyed and reloaded. This is necessary when a majority of matching replies is not achieved or the Compute Platform is slow or unresponsive. In the case of stateful applications, the state saved in the PLM-C BRAMs is first checkpointed and saved in the on-chip SRAM before triggering the Tileloader, to ensure no state progress is lost. The \textit{Tileloader} reloads the Compute Platform following a predefined policy as follows:
    \begin{enumerate}
        \item Refresh/Diversify: reloads the exact (Refresh) or diverse (Diversify) softcore version to the PL. Refresh is convenient for transient faults or attacks while Diversify boosts the Compute Platform resilience through minimizing common-mode failures.
        \item Replace/Relocate: reloads the softcore in the same or different PL partition location. This precludes potential issues in the PL fabric location and changes the communication routes in the case of AXI issues/attacks. Relocation, however, assumes a block for the relocation of the partial bitstream into a new location and its interface has been taken into account at design time during floorplanning.
        \item Scale out/in: changes the number of simultaneous tiles adapting to severity levels, i.e., as $f$ in $2f+1$ changes. This is bounded by the \textit{min-tiles} and \textit{max-tiles} parameters in configuration (see Table~\ref{tab:config}).
        \item Reactive/Proactive: invokes Tileloader in reaction to a fault raised by \proto or in a periodic fashion regardless of faults, seeking proactive resilience (especially useful against \textit{Advanced Persistent Threats}).
    \end{enumerate}
    \item The \textit{Tileloader} notifies $C$ with \textit{Ready} status.
    \item $C$ sets a timer and waits for \textit{Ready} response from the corresponding rejuvenated tiles.
    \item If $C$ received the expected number of \textit{Ready} messages from the tiles before the timer's expiry, it sets its $\mathbf{status}\leftarrow Ready$, transfers the checkpointed state back to PLM-C, and starts accepting application requests. Otherwise, the Controller launches the \textit{Rejuvenation} phase in \textit{partial-mode} if minority of Tiles are faulty or slow, or in full-mode otherwise.
\end{enumerate}

\section{Evaluation}
\label{sec:evaluation}

The proposed framework was evaluated on a Xilinx Zynq Ultrascale+ ZCU102 FPGA board, running at 100MHz. The choice of frequency has the goal of fair comparison, as the references against which we compare Samsara run also at 100MHz. We instantiated 3 reconfigurable partitions (RP) in the PL (i.e., 3 tiles), each with 2 possible reconfigurable modules (RM). RMs refer to the possible partial bitstreams that fit into the RPs which are dedicated to tile implementation. The choice of the type of IPs to run on the PL, i.e., what tiles contain, is application dependent and not a property of Samsara. RMs can only be loaded into a full bitstream with the RPs designed to contain them. However, by triggering full Rejuvenation and replacing the whole PL, it is possible to install a new full bitstream that can interface with different types of RMs. Namely, a bitstream can be designed to have more generic RP boundaries which interface more easily with different types of tiles. In order to evaluate the most complex and costly case possible for Samsara, we designed the tiles to be softcores (Xilinx's MicroBlazes in standard configuration). Simpler cases include cryptographic or machine learning accelerators, or any other type of module. As a proof of concept, the Controller is simulated in the PL and not implemented as an ASIC (as it is would require manufacturing and be costly to experiment with over several iterations). It is then instead implemented as software running on a PL softcore. Tiles' cores and other components in the PL (e.g., memory controllers, timers, etc) are connected using AXI4-Lite. The Bootloader and Tileloader are run in one of the FPGA's PS ARM cores.

Table~\ref{fig:eval_protocols} details the architectures/protocols evaluated in this paper. We evaluate two fault models for Samsara, (1) one where we assume the MPSoC NoC and AXI buses to be correct and thus do not use message hashes; and (2) one where we acknowledge network attacks (as described in the threat model in Section~\ref{sec:threat}) and, thus, use hashes (\emph{Samsara HW\_H} in Table~\ref{fig:eval_protocols}). Both scenario (1) and (2) use 3 partitions as described above. We compare these implementations with a baseline of having just one \emph{Tile}, in this case, one softcore (\emph{SC} in Table~\ref{fig:eval_protocols}), with a regular triple-modular redundant (TMR) architecture (with and without hashing), the \emph{iBFT} protocol from~\cite{pinto2022architectural} and a shared-memory-based implementation of \emph{MinBFT}~\cite{MinBFT2013} akin to that described in~\cite{pinto2022architectural}. Using 3 \emph{Tiles} is the minimum to have majority (i.e., if $f = 1$, then $n = 2f + 1 = 3$) and allows better comparison to basic redundancy (TMR) and~\cite{pinto2022architectural}, since it uses 3 replicas as well. The comparison with \emph{iBFT} is relevant given it too implements a low-level agreement protocol like \proto{}. The other protocol presented in~\cite{pinto2022architectural}, \emph{Midir}~\cite{gouveia2022behind}, is slower than \emph{iBFT} and, therefore, is not as suitable for comparison. Samsara's version with no hashes also helps to compare with \emph{iBFT}, which assumes a correct network. \emph{MinBFT} is the most well-know state-of-the-art BFT protocol that implements architectural hybridization~\cite{verissimo2003uncertainty} and, as such, was a good target for comparison.

\begin{table}
  \caption{Protocol explanations. HW = hardware and SW = software.}
  \label{fig:eval_protocols}
  \begin{center}
    \begin{tabular}{|l|l|}
      \hline
      SC & Single Core (Baseline)\\\hline
      TMR & 3 Cores\\\hline
      TMR HW\_H & 3 Cores with HW Hashing (SHA-256)\\\hline
      H-Quorum & 3 Cores\\\hline
      H-Quorum HW\_H & 3 Cores with HW Hashing (SHA-256)\\\hline
      iBFT~\cite{pinto2022architectural} & 3 Cores\\\hline
      MinBFT~\cite{MinBFT2013} SW\_H & 3 Cores with SW Hashing (SHA-256)\\\hline
    \end{tabular}
  \end{center}
\end{table}

The protocols are compared in terms of agreement latency, reconfiguration latency, latency footprint, area usage, power consumption and communication steps. Area usage and power consumption are SWaP (space, weight and power) metrics, which are relevant in the construction of on-chip, often resource restricted systems like cyber-physical systems (CPS) and IoT.

\subsection{Latency}
Latency is measured with a combination of an AXI Timer and an AXI Interrupt Controller module in the PL, and a Xilinx timer API that outputs the cycle count.

\paragraph{\textbf{Protocol Latency}}

\begin{figure}[t]
\centering
\includegraphics[width=.9\columnwidth]{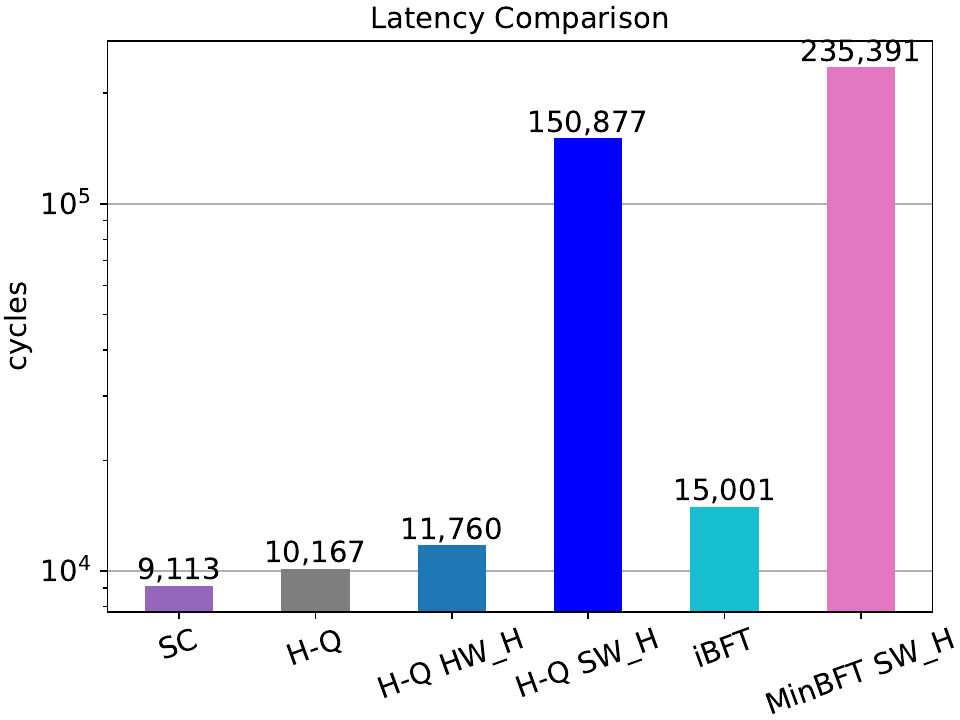}
\caption{Latency comparison of Samsara's H-Quorum (H-Q in the graph) with and without hashing (null operation) against a baseline (single core), iBFT and MinBFT. The graph is presented in logarithmic scale, but the values seen atop the bars are the true cycle count.}
\label{fig:latency}
\end{figure}

Fig.~\ref{fig:latency} compares Samsara's \proto{} (H-Q in the figure) with 256-bit messages (with and without hashing) with our baseline (\emph{SC}), \emph{iBFT} and the shared-memory implementation of \emph{MinBFT}, in logarithmic scale. It is observable that the use of 3 agreeing \emph{Tiles} as opposed to 1 only has a latency overhead of 1054 cycles (i.e., 10.54 ms, considering a core running at 100MHz). Additionally, \proto{} with hardware-based hashing is still 3241 cycles faster (and 4834 cycles faster without hashing) than \emph{iBFT} (no hashing), meaning it is 21.6\% (and 32.2\% respectively) faster than \emph{iBFT}. In Samsara, we use an SHA-256 PL module that we connected with each core \emph{Tile} (so 3 SHA-256 modules, one for each \emph{Tile}) to perform the hashing of the messages. The lower cycle count in comparison with \emph{iBFT} is expected, as \emph{iBFT} has more communication steps and number of exchanged messages, as seen in Table~\ref{fig:comm_steps}. All measurements are done for a null operation, meaning the depicted latency represent the protocol latency (i.e., message exchange, which in this case is done through reads and writes from/to BRAM) and does not include request execution latency.

To fairly compare with MinBFT which has hashing implemented in software, we present the results for an \proto{} implementation that uses SHA-256 in software rather than hardware, \proto{} SW\_H. As can be seen, \proto{} performs faster due to having less protocol steps.

In the FPGA realm, with hardware accelerators significantly reducing latency (e.g, SHA-256 in hardware takes on average 1593 cycles, while in software, running on a MicroBlaze core, it takes 128643 cycles), the biggest sources of latency are clock domain crossing and the nature of the bus protocol implementation (for simplicity, we used Xilinx's implementation of AXI4-Lite, which is slower than AXI4-Full and does not allow burst access) for memory access. With more complex implementations of the AXI4 bus and higher clock frequencies, better performance can be achieved.

\begin{table}
  \caption{Complexity estimation of TMR, Samsara (H-Quorum), iBFT and MinBFT.}
  \label{fig:comm_steps}
  \begin{center}
    \begin{tabular}{|l|l|l|l|}
      \hline
      Protocol & Communication Steps & Number of Messages \\\hline
      TMR & 1 & O(n) \\\hline
      H-Quorum & 2 & O(n) \\\hline
      iBFT & 5 & O(n\textsuperscript{2}) \\\hline
      MinBFT & 4 & O(n\textsuperscript{2}) \\\hline
    \end{tabular}
  \end{center}
\end{table}

\paragraph{\textbf{Protocol Footprint}}

\begin{figure}[t]
\centering
\includegraphics[width=.9\columnwidth]{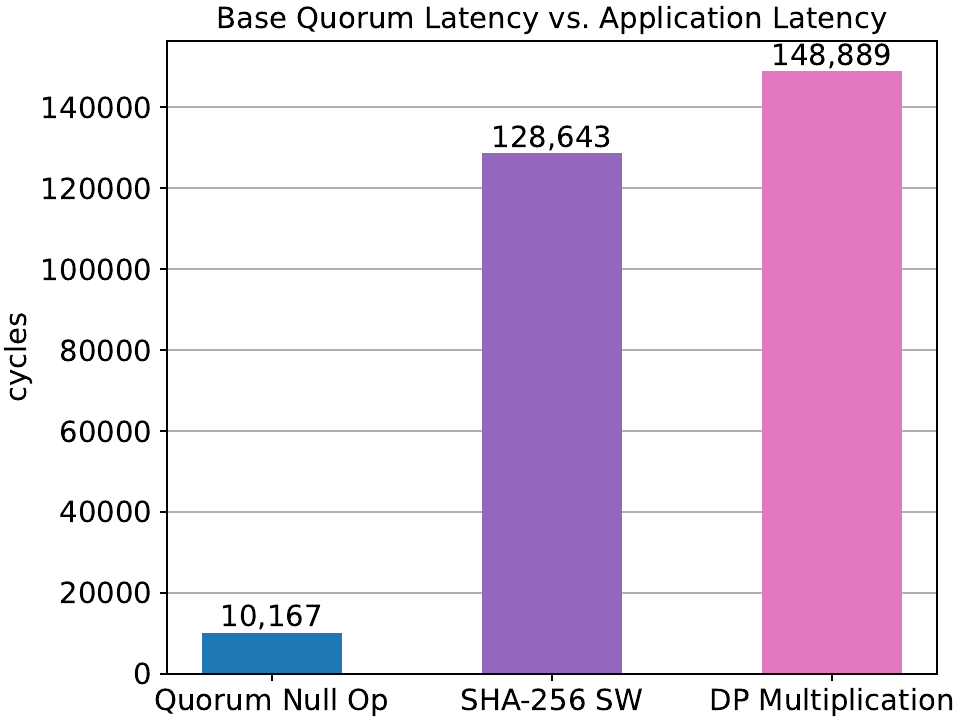}
\caption{Influence of application complexity on agreement latency. The null operation represents the latency of the Quorum protocol itself. DP Multiplication means double percision multiplication.}
\label{fig:quorum_negligible}
\end{figure}

Fig.~\ref{fig:quorum_negligible} shows how \proto{}'s latency (null operation) becomes more negligible as application complexity scales. We ran SHA-256 on a 256-bit message in software (i.e., running on the MicroBlaze core and not the hardware accelerator used in Samsara) and performed complex double-precision multiplication on a one-dimensional array of very small size (10) for comparison; and observed that running the \proto{} agreement protocol brings no significant overhead to operation execution. Naturally, these operations can be run on dedicated accelerators, which would replace cores in the \emph{Tiles} and have lower latency. Namely, \proto{} itself will have better performance with the envisioned ASIC controller, as it will execute as hardware logic. However, given we are emulating the \emph{Controller} in software running on a PL core, it is fair to compare against software implementations of the aforementioned operations. Even with, e.g., a hardware implementation of SHA-256 (which we use in Samsara, and takes 1593 cycles for a 256-bit input message), \proto{} is, not only still fast, but does not incur great overhead based on replication or agreement, as previously seen in Fig.~\ref{fig:latency} which shows single core execution as taking only 1054 cycles less.

\paragraph{\textbf{Reconfiguration Latency}}

\begin{figure}[t]
\centering
\includegraphics[width=.9\columnwidth]{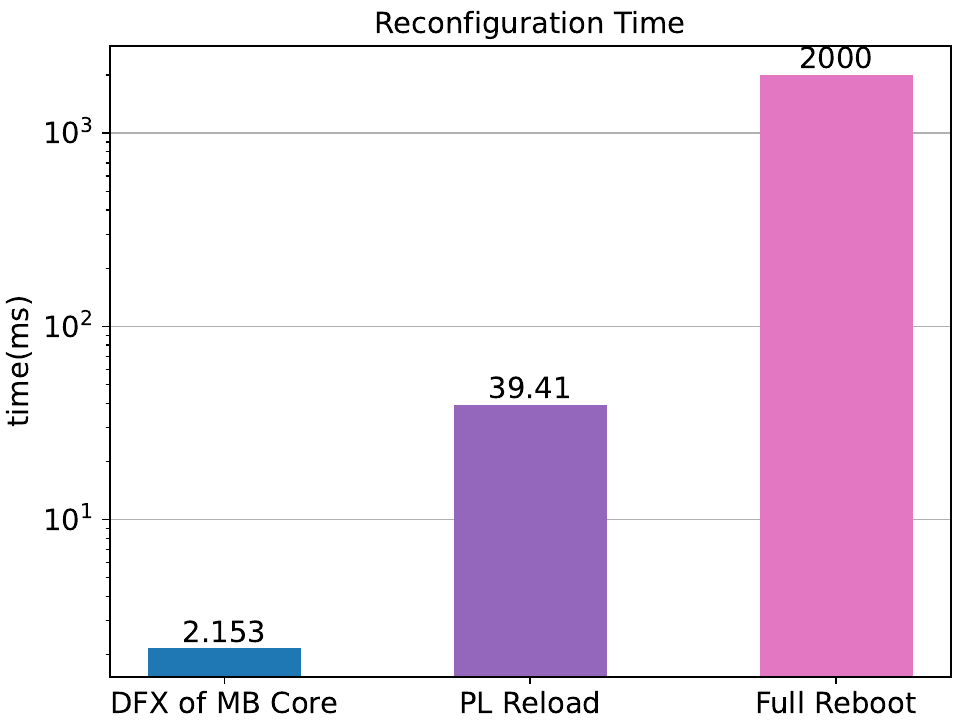}
\caption{Reconfiguration time in milliseconds (ms) of DFX (dynamic partial reconfiguration) vs. total PL reboot vs. total board reboot, in logarithmic scale.}
\label{fig:reconf_time}
\end{figure}

We then analyzed reconfiguration time, measuring the time to partially reconfigure the PL (i.e., replacing a \emph{Tile}, in this case, a MicroBlaze \emph{Tile}) against the time to fully reconfigure the whole PL, and a full platform reboot (i.e., including running the Bootloader again and initializing the PS), as seen in Fig.~\ref{fig:reconf_time}. The partial and full PL reconfiguration are measured by the PS's ARM core, while the full platform reboot is measured with the help of the Xilinx application development tool, Vitis which logs the time elapsed from the beginning of the reboot process until it has successfully finished, excluding the time it takes to download the bitstreams and \texttt{.elf} files to the board through JTAG. Given that, by rebooting the whole board we do not have any cores available to perform the latency measurement, we could not check the time elapsed by booting the board from the FPGA's SD card (as done in the PL reconfiguration measurements), and therefore we required the time measurement provided by the tool. As such, for the full reboot we programmed the board via JTAG. We evaluated this process through JTAG, which gave us a very precise measurement from the tool. Note that the full reboot latency presented in Fig.~\ref{fig:reconf_time} does not account for applications loaded into the PS, besides the Bootloader and Tileloader.

\begin{figure}[t]
\centering
\includegraphics[width=\columnwidth]{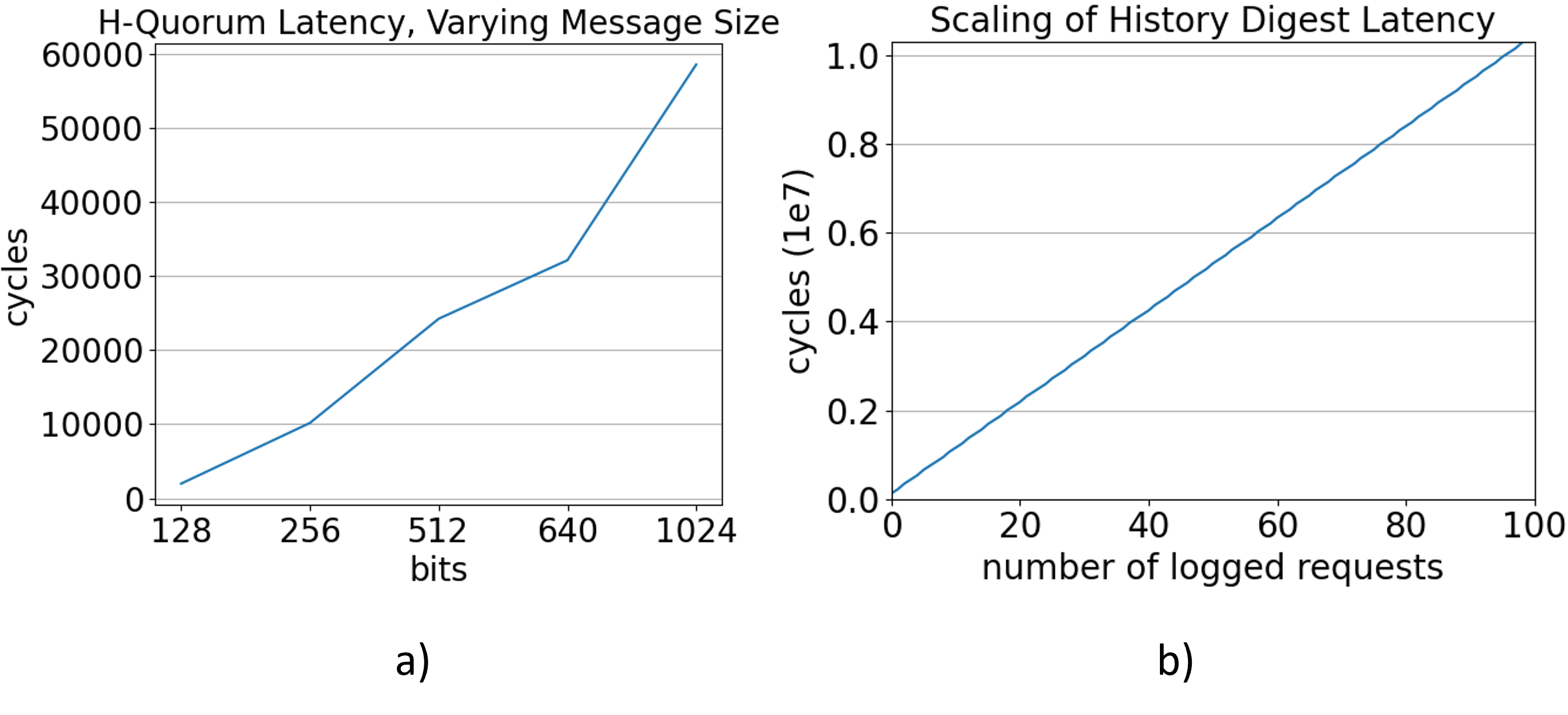}
\caption{a) Latency of \proto{} with varying message sizes for null operation. b) Scaling of history digest latency (each round), according to the number of logged requests (up to a checkpoint of 100) for stateful applications.}
\label{fig:msg_size_digest}
\end{figure}

As can be seen, there is a significant decrease in the time required to reconfigure a \emph{Tile} versus the full PL, which results from replacing only a small part of the architecture. One MicroBlaze core represents only 4.9\% of the Samsara design, i.e., of the full base bitstream. The full reboot, on the other hand, requires reset and initialization of the PS, running the FSBL and programming of the PL. The presented results were obtained with un-encrypted bitstreams, meaning they do not account for bitstream decryption, however, the reduced latency gained from PR would still hold. For stateful applications that need a local state copy, state hashing (digest) and transfer latency depends on message and checkpoint size. Fig.~\ref{fig:msg_size_digest} a) depicts \proto{}'s latency as a function of varying message sizes, again using a null operation. As expected, latency grows linearly with the rising memory accesses. Our AXI4-Lite interface performs writes of 32 bits, with number of accesses scaling linearly with message length. Other bus implementation options include burst or streaming. Similarly, Fig.~\ref{fig:msg_size_digest} b) shows latency hashing latency (computed in software) with a growing checkpoint size, up to 100.

\subsection{Resource Usage and Power Consumption}

\begin{figure}[t]
\centering
\includegraphics[width=\columnwidth]{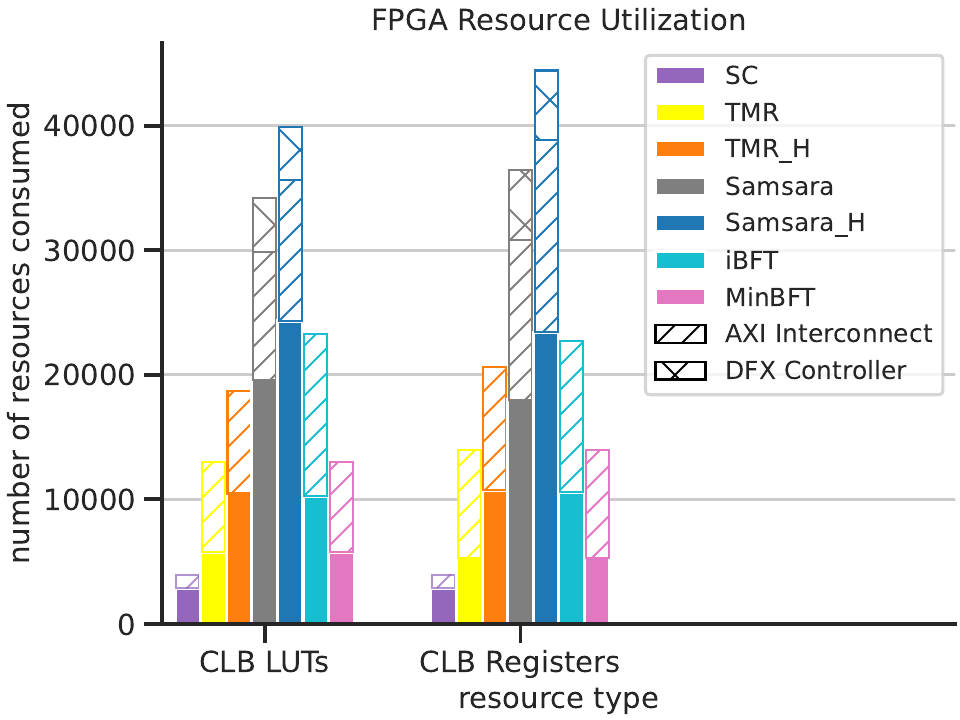}
\caption{FPGA resource utilization.}
\label{fig:fpga_resource_utilization}
\end{figure}

Next we look at resource usage, as seen in Fig.~\ref{fig:fpga_resource_utilization}. The metrics presented in the graph are LUTs (Look-Up Tables) and registers which are the basic unit of area measurement on the FPGA fabric. We divided each architecture into 3 categories: base architecture (\emph{Tiles}, memory controllers, timers, interrupts, etc), AXI Interconnect and DFX~\footnote{DFX stands for Dynamic Function eXchange and refers to the Xilinx functionality used to perform dynamic partial reconfiguration at runtime.} Controller (not to be mistaken with Samsara's \emph{Controller}). The DFX Controller in the PL manages the low-level loading of hardware bitstreams from memory. The separation of the AXI from the rest of the architecture has the intent to show the large occupation of the bus in comparison to the rest of the design, it is still, however, part of the PL along with the DFX Controller. The AXI Interconnect module connect several PL modules and thus multiplexes accesses among them, consuming a large number of LUTs and registers.
This is the reason why AXI increases significantly from using a single \emph{Tile} (\emph{SC}) to using 3 (Samsara). For simplicity and easy comparison with the selected state of the art, we implemented Samsara with main AXI Interconnect for connecting most modules and used the AXI4-Lite interface. Separating it into multiple smaller ones can lower resource consumption as there is less multiplexing, while using AXI4-Full can bring lower latency. DFX is naturally only present in Samsara, as it is the only solution presented here that does dynamic partial reconfiguration, i.e., swapping \emph{Tile} contents at runtime. This represents a trade-off of between a slightly higher area and resource usage, and the flexibility and speed of hardware reconfiguration. In the evaluated board (ZCU102), the utilization for Samsara still represents only 12.68\% of available LUTs and 6.65\% of available registers, with 3.24\%/1.92\% (respectively) for the DDR4 memory where bitstreams were stored, 2.38\%/1.63\% for the AXI Interconnect, and 1.59\%/1.02\% for the DFX Controller.

Table~\ref{fig:power} shows the predicted power consumption for all evaluated architectures. This power analysis is taken from the implemented netlist outputted by the Xilinx design, synthesis and implementation tool Vivado. It represents activity derived from constraint files, simulation files and vectorless analysis and is provided as expectations, not physically tested results.

As expected, Samsara consumes approximately 1.297 W more than the average of the other architectures, which comes from the usage of DFX. Nevertheless, in Samsara, 2.744 W out of the 5.064 W come from the PS and not the PL where the \emph{Tiles} reside. It is also noticeable that the usage of 3 \emph{Tiles} (TMR and Samsara) versus 1 \emph{Tile} leads only to an increase of 0.146 W, meaning that replication or multiple \emph{Tiles} participating in \proto{} does not massively affect power consumption.

\begin{table}
  \caption{Total on-chip power consumption in watts (W).}
    \label{fig:power}
  \begin{center}
    \begin{tabular}{|l|l|}
      \hline
      SC & 3.623 W\\\hline
      TMR & 3.769 W\\\hline
      TMR HW\_H & 3.813 W\\\hline
      Samsara & 5.064 W\\\hline
      Samsara HW\_H & 5.108 W\\\hline
      iBFT & 3.973 W\\\hline
      MinBFT & 3.769 W\\\hline
    \end{tabular}
  \end{center}
\end{table}


\section{Related Works}
\label{sec:related-work}

TMR has been used in the realm of critical embedded systems, e.g., in the primary flight computers of
Boeing 777's fly-by-wire (FBW) system~\cite{yeh1996triple}. Similarly, passive redundancy can also be seen in Airbus' dependability-oriented approach to FBW~\cite{traverse2004airbus}. The concept was extended to multi-phase tightly-synchronous message-passing protocols in the CPS domain~\cite{mancini1986modular, kopetz2003time}. Unlike TMR, Byzantine fault tolerant (BFT) SMR algorithms~\cite{BFT1982, PBFT1999} aim to tolerate both accidental and malicious faults, by reaching consensus with $|Q|=2f+1$ out of $n=3f+1$ replicas. Architectural hybridization~\cite{Verissimo2006} proposes an additional trusted-trustworthy component to further reduce the size of $n$ and $Q$ to $2f+1$ and $f+1$, respectively. This technique has been used in protocols such as MinBFT and CheapBFT~\cite{MinBFT2013,kapitza2012cheapbft}. BFT algorithms have traditionally been implemented on distributed systems and have seen little work in emerging MPSoC critical systems like CPS and IoT, due to their added latency and replication costs. \emph{Midir} presents an architecture and an on-chip BFT-like protocol for improving the safety and resilience of low-level software running on MPSoCs as well as their access control mechanisms. It does so through minimalist hardware logic, T2H2, that provides secure voting on critical operations and access control. Similarly, \textit{iBFT}~\cite{pinto2022architectural} is an efficient consensus algorithm by leveraging shared memory. Nevertheless, neither of these works allow the flexilibity or accelerator reconfiguration not dynamic rejuvenation. In~\cite{avramenko2019rtos}, a partitioning technique enabling the use of COTS NoC-based MPSoC for mixed criticality systems is proposed, however, it is intended to have a purely software implementation as a module of a real-time operating system. Another work~\cite{pagonis2023increasing} investigates and evaluates fault-tolerance techniques for the UltraScale+ MPSoC FPGA, but it targets only accidental faults in the form of Single-Event Upsets (SEUs). Contention on shared resources in the context of MPSoC-based safety critical applications is explored in~\cite{tabish2021next}. Alcaide et al.~\cite{alcaide2022unboxing} develop safety measures in the PL of COTS MPSoCs, but do not consider PL-side faults and intrusions. In~\cite{nicholas2021secure}, a secure framework to implement logic-locking, extended with secure boot for Xilinx FPGAs is presented. Furkan~\cite{Turan2021} provides a survey on secure FPGA configuration and security of FPGA modules.

\section{Conclusion}
\label{sec:conclusion}

We introduced \textit{Samsara}, the first secure and resilient platform for programmable hardware. Our work shows that leveraging the hardware properties of programmable hardware, like FPGA and GPU, paves the way to design lightweight and low latency BFT variants, such as \textit{H-Quorum}. Interestingly, we show that hardware rejuvenation is also possible at a negligible latency overhead. To improve independence of failures, Samsara supports the rejuvenation to diverse implementations from a pool of versions. This is possible either by simple reconfiguration tweaks or via using off-the-shelf implementations. In particular, typical compute IP implementations, e.g., SHA-256, are available as open source. Samsara, however, imposes an additional resource utilization overhead over non-replicated systems, which we believe is a reasonable price for security and resilience. This is not as critical as replicated computers since a programmable fabric might often be under-utlized.

\newpage

\bibliographystyle{IEEEtran}
\bibliography{IEEEabrv,main}

\begin{thebibliography}{10}
\providecommand{\url}[1]{#1}
\csname url@samestyle\endcsname
\providecommand{\newblock}{\relax}
\providecommand{\bibinfo}[2]{#2}
\providecommand{\BIBentrySTDinterwordspacing}{\spaceskip=0pt\relax}
\providecommand{\BIBentryALTinterwordstretchfactor}{4}
\providecommand{\BIBentryALTinterwordspacing}{\spaceskip=\fontdimen2\font plus
\BIBentryALTinterwordstretchfactor\fontdimen3\font minus \fontdimen4\font\relax}
\providecommand{\BIBforeignlanguage}[2]{{%
\expandafter\ifx\csname l@#1\endcsname\relax
\typeout{** WARNING: IEEEtran.bst: No hyphenation pattern has been}%
\typeout{** loaded for the language `#1'. Using the pattern for}%
\typeout{** the default language instead.}%
\else
\language=\csname l@#1\endcsname
\fi
#2}}
\providecommand{\BIBdecl}{\relax}
\BIBdecl

\bibitem{Safarulla2014}
I.~M. Safarulla and K.~Manilal, ``Design of soft error tolerance technique for {FPGA} based soft core processors,'' in \emph{IEEE International Conference on Advanced Communications, Control and Computing Technologies}, 2014, pp. 1036--1040.

\bibitem{Shashidhara2020}
B.~Shashidhara, S.~Jadhav, and Y.~S. Kim, ``Reconfigurable {F}ault {T}olerant {P}rocessor on a {SRAM} based {FPGA},'' in \emph{IEEE International Conference on Electro Information Technology (EIT)}, 2020, pp. 151--154.

\bibitem{Feng2022}
H.~Feng, W.~Li, L.~Chen, S.~Wang, J.~Zhou, C.~Tian, and Y.~Zhang, ``Precise {F}ault {I}njection and {F}ault {L}ocation {S}ystem for {SRAM}-based {FPGA}s,'' in \emph{IEEE 10th Joint International Information Technology and Artificial Intelligence Conference (ITAIC)}, vol.~10, 2022, pp. 2371--2376.

\bibitem{SGX2016}
\BIBentryALTinterwordspacing
V.~Costan and S.~Devadas, ``Intel {SGX} explained,'' \emph{Tech. rep., Massachusetts Institute of Technology}, 2016. [Online]. Available: \url{https://eprint.iacr.org/2016/086.pdf}
\BIBentrySTDinterwordspacing

\bibitem{ARMTrustZone}
\BIBentryALTinterwordspacing
ARM, ``{ARMT}rust{Z}one,'' \emph{ARM}, Accessed on August 2022. [Online]. Available: \url{https://www.arm.com/products/security-on-arm/trustzone}
\BIBentrySTDinterwordspacing

\bibitem{kinney2006trusted}
S.~L. Kinney, \emph{Trusted platform module basics: using TPM in embedded systems}.\hskip 1em plus 0.5em minus 0.4em\relax Elsevier, 2006.

\bibitem{sabt2015trusted}
M.~Sabt, M.~Achemlal, and A.~Bouabdallah, ``Trusted execution environment: what it is, and what it is not,'' in \emph{IEEE Trustcom/BigDataSE/ISPA}, vol.~1.\hskip 1em plus 0.5em minus 0.4em\relax IEEE, 2015, pp. 57--64.

\bibitem{Distler2016}
T.~Distler, C.~Cachin, and R.~Kapitza, ``Resource-{E}fficient {B}yzantine {F}ault {T}olerance,'' \emph{IEEE Transactions on Computers}, vol.~65, no.~9, pp. 2807--2819, 2016.

\bibitem{hw-assisted-2018scalable}
J.~Liu, W.~Li, G.~O. Karame, and N.~Asokan, ``Scalable byzantine consensus via hardware-assisted secret sharing,'' \emph{IEEE Transactions on Computers}, vol.~68, no.~1, pp. 139--151, 2018.

\bibitem{MinBFT2013}
G.~S. Veronese, M.~Correia, A.~N. Bessani, L.~C. Lung, and P.~Verissimo, ``{E}fficient {B}yzantine {F}ault-{T}olerance,'' \emph{IEEE Transactions on Computers}, vol.~62, no.~1, pp. 16--30, 2013.

\bibitem{ZynqMPSoC}
\BIBentryALTinterwordspacing
AMD, ``Zynq {U}ltrascale+ {D}evice {T}echnical {R}eference {M}anual,'' \emph{Xilinx}, 2020. [Online]. Available: \url{https://docs.xilinx.com/v/u/en-US/ug1085-zynq-ultrascale-trm}
\BIBentrySTDinterwordspacing

\bibitem{shoker2023path}
A.~Shoker, P.~Esteves-Verissimo, and M.~V{\"o}lp, ``The {P}ath to {F}ault-and {I}ntrusion-{R}esilient {M}anycore {S}ystems on a {C}hip,'' in \emph{53rd Annual IEEE/IFIP International Conference on Dependable Systems and Networks-Supplemental Volume (DSN-S)}.\hskip 1em plus 0.5em minus 0.4em\relax IEEE, 2023, pp. 157--162.

\bibitem{karandikar2018firesim}
S.~Karandikar, H.~Mao, D.~Kim, D.~Biancolin, A.~Amid, D.~Lee, N.~Pemberton, E.~Amaro, C.~Schmidt, A.~Chopra \emph{et~al.}, ``Fire{S}im: {FPGA}-accelerated cycle-exact scale-out system simulation in the public cloud,'' in \emph{ACM/IEEE 45th Annual International Symposium on Computer Architecture (ISCA)}.\hskip 1em plus 0.5em minus 0.4em\relax IEEE, 2018, pp. 29--42.

\bibitem{esp2020}
P.~Mantovani, D.~Giri, G.~Di~Guglielmo, L.~Piccolboni, J.~Zuckerman, E.~G. Cota, M.~Petracca, C.~Pilato, and L.~P. Carloni, ``Agile {S}o{C} {D}evelopment with {O}pen {ESP} : {I}nvited {P}aper,'' in \emph{IEEE/ACM International Conference On Computer Aided Design (ICCAD)}, 2020, pp. 1--9.

\bibitem{singh2000morphosys}
H.~Singh, M.-H. Lee, G.~Lu, F.~J. Kurdahi, N.~Bagherzadeh, and E.~M. Chaves~Filho, ``{MorphoSys}: an integrated reconfigurable system for data-parallel and computation-intensive applications,'' \emph{IEEE Transactions on Computers}, vol.~49, no.~5, pp. 465--481, 2000.

\bibitem{boutros2021fpga}
A.~Boutros and V.~Betz, ``{FPGA} architecture: Principles and progression,'' \emph{IEEE Circuits and Systems Magazine}, vol.~21, no.~2, pp. 4--29, 2021.

\bibitem{mal2014hardware}
S.~Mal-Sarkar, A.~Krishna, A.~Ghosh, and S.~Bhunia, ``Hardware {T}rojan {A}ttacks in {FPGA} {D}evices: {T}hreat {A}nalysis and {E}ffective {C}ountermeasures,'' in \emph{Proceedings of the 24th Edition of the Great Lakes Symposium on VLSI}, 2014, pp. 287--292.

\bibitem{duncan2019fpga}
A.~Duncan, F.~Rahman, A.~Lukefahr, F.~Farahmandi, and M.~Tehranipoor, ``{FPGA} bitstream security: a day in the life,'' in \emph{IEEE International Test Conference (ITC)}.\hskip 1em plus 0.5em minus 0.4em\relax IEEE, 2019, pp. 1--10.

\bibitem{ender2020unpatchable}
M.~Ender, A.~Moradi, and C.~Paar, ``The unpatchable silicon: a full break of the bitstream encryption of {X}ilinx 7-series {FPGA}s,'' in \emph{29th USENIX Security Symposium (USENIX Security 20)}, 2020, pp. 1803--1819.

\bibitem{ender2022cautionary}
M.~Ender, G.~Leander, A.~Moradi, and C.~Paar, ``A cautionary note on protecting {X}ilinx’s {U}ltra{S}cale+ bitstream encryption and authentication engine,'' in \emph{IEEE 30th Annual International Symposium on Field-Programmable Custom Computing Machines (FCCM)}.\hskip 1em plus 0.5em minus 0.4em\relax IEEE, 2022, pp. 1--9.

\bibitem{BFT1982}
\BIBentryALTinterwordspacing
L.~Lamport, R.~Shostak, and M.~Pease, ``The byzantine generals problem,'' \emph{ACM Trans. Program. Lang. Syst.}, vol.~4, no.~3, p. 382–401, jul 1982. [Online]. Available: \url{https://doi.org/10.1145/357172.357176}
\BIBentrySTDinterwordspacing

\bibitem{BitStreamEncryption}
\BIBentryALTinterwordspacing
AMD, ``Using {E}ncryption and {A}uthentication to {S}ecure an {U}ltra{S}cale/{U}ltra{S}cale+ {FPGA} {B}itstream,'' \emph{Xilinx}, 2022. [Online]. Available: \url{https://www.xilinx.com/content/dam/xilinx/support/documents/application_notes/xapp1267-encryp-efuse-program.pdf}
\BIBentrySTDinterwordspacing

\bibitem{zhang2014secure}
T.~Zhang and R.~B. Lee, ``Secure cache modeling for measuring side-channel leakage,'' \emph{Technical Report, Princeton University}, 2014.

\bibitem{avramenko2019rtos}
S.~Avramenko and M.~Violante, ``{RTOS} solution for {N}o{C}-based {COTS} {MPS}o{C} usage in mixed-criticality systems,'' \emph{Journal of Electronic Testing}, vol.~35, pp. 29--44, 2019.

\bibitem{charles2021survey}
S.~Charles and P.~Mishra, ``A survey of network-on-chip security attacks and countermeasures,'' \emph{ACM Computing Surveys (CSUR)}, vol.~54, no.~5, pp. 1--36, 2021.

\bibitem{lyons1962use}
R.~E. Lyons and W.~Vanderkulk, ``The use of triple-modular redundancy to improve computer reliability,'' \emph{IBM journal of research and development}, vol.~6, no.~2, pp. 200--209, 1962.

\bibitem{fischer1985impossibility}
M.~J. Fischer, N.~A. Lynch, and M.~S. Paterson, ``Impossibility of distributed consensus with one faulty process,'' \emph{Journal of the ACM (JACM)}, vol.~32, no.~2, pp. 374--382, 1985.

\bibitem{makni2016comparison}
M.~Makni, M.~Baklouti, S.~Niar, M.~W. Jmal, and M.~Abid, ``A comparison and performance evaluation of {FPGA} soft-cores for embedded multi-core systems,'' in \emph{11th International Design \& Test Symposium (IDT)}.\hskip 1em plus 0.5em minus 0.4em\relax IEEE, 2016, pp. 154--159.

\bibitem{waterman2014risc}
A.~Waterman, Y.~Lee, D.~Patterson, K.~Asanovic, V.~I.~U. level Isa, A.~Waterman, Y.~Lee, and D.~Patterson, ``The {RISC}-{V} instruction set manual,'' \emph{Volume I: User-Level ISA’, version}, vol.~2, pp. 1--79, 2014.

\bibitem{marques2021lock}
I.~Marques, C.~Rodrigues, A.~Tavares, S.~Pinto, and T.~Gomes, ``Lock-{V}: {A} heterogeneous fault tolerance architecture based on {ARM} and {RISC-V},'' \emph{Microelectronics Reliability}, vol. 120, p. 114120, 2021.

\bibitem{guerraoui2010next}
R.~Guerraoui, N.~Kne{\v{z}}evi{\'c}, V.~Qu{\'e}ma, and M.~Vukoli{\'c}, ``The next 700 {BFT} protocols,'' in \emph{Proceedings of the 5th European conference on Computer systems}, 2010, pp. 363--376.

\bibitem{adapt:2015}
J.-P. Bahsoun, R.~Guerraoui, and A.~Shoker, ``Making {BFT} protocols really adaptive,'' in \emph{IEEE International Parallel and Distributed Processing Symposium}.\hskip 1em plus 0.5em minus 0.4em\relax IEEE, 2015, pp. 904--913.

\bibitem{zcu102}
AMD, \emph{ZCU102 Evaluation Board User Guide (UG1182)}, 2023.

\bibitem{pinto2022architectural}
I.~Pinto~Gouveia, ``Architectural support for hypervisor-level intrusion tolerance in mpsocs,'' Ph.D. dissertation, University of Luxembourg, Esch-sur-Alzette, Luxembourg, 2022.

\bibitem{zynq7000}
AMD, \emph{Zynq 7000 SoC Technical Reference Manual (UG585), version 1.12.2}, 2018.

\bibitem{tpm}
T.~C. Group, \emph{TPM Main, Part 1 Design Principles, Specification Version 1.2, Revision 116}, 2011.

\bibitem{sharma2007onboard}
A.~K. Sharma, ``Onboard credentials: {H}ardware assisted secure storage of credentials,'' Master's thesis, Helsinki University of Technology, 2007.

\bibitem{10.1145/3503222.3507742}
\BIBentryALTinterwordspacing
D.~Cock, A.~Ramdas, D.~Schwyn, M.~Giardino, A.~Turowski, Z.~He, N.~Hossle, D.~Korolija, M.~Licciardello, K.~Martsenko, R.~Achermann, G.~Alonso, and T.~Roscoe, ``Enzian: {A}n {O}pen, {G}eneral, {CPU}/{FPGA} {P}latform for {S}ystems {S}oftware {R}esearch,'' in \emph{Proceedings of the 27th ACM International Conference on Architectural Support for Programming Languages and Operating Systems}, ser. ASPLOS 2022.\hskip 1em plus 0.5em minus 0.4em\relax New York, NY, USA: Association for Computing Machinery, 2022, p. 434–451. [Online]. Available: \url{https://doi.org/10.1145/3503222.3507742}
\BIBentrySTDinterwordspacing

\bibitem{waingold1997baring}
E.~Waingold, M.~Taylor, D.~Srikrishna, V.~Sarkar, W.~Lee, V.~Lee, J.~Kim, M.~Frank, P.~Finch, R.~Barua \emph{et~al.}, ``Baring it all to software: Raw machines,'' \emph{Computer}, vol.~30, no.~9, pp. 86--93, 1997.

\bibitem{liu2019survey}
L.~Liu, J.~Zhu, Z.~Li, Y.~Lu, Y.~Deng, J.~Han, S.~Yin, and S.~Wei, ``A survey of coarse-grained reconfigurable architecture and design: {T}axonomy, challenges, and applications,'' \emph{ACM Computing Surveys (CSUR)}, vol.~52, no.~6, pp. 1--39, 2019.

\bibitem{sec_pcie}
I.~Corporation, \emph{PCI Express® Device Security Enhancements}, 2018.

\bibitem{vipin2018fpga}
K.~Vipin and S.~A. Fahmy, ``{FPGA} dynamic and partial reconfiguration: A survey of architectures, methods, and applications,'' \emph{ACM Computing Surveys}, vol.~51, no.~4, pp. 1--39, 2018.

\bibitem{chakraborty2013hardware}
R.~S. Chakraborty, I.~Saha, A.~Palchaudhuri, and G.~K. Naik, ``Hardware {T}rojan insertion by direct modification of {FPGA} configuration bitstream,'' \emph{IEEE Design \& Test}, vol.~30, no.~2, pp. 45--54, 2013.

\bibitem{swierczynski2017bitstream}
P.~Swierczynski, G.~T. Becker, A.~Moradi, and C.~Paar, ``Bitstream fault injections ({B}i{F}i)--automated fault attacks against {SRAM}-based {FPGA}s,'' \emph{IEEE Transactions on Computers}, vol.~67, no.~3, pp. 348--360, 2017.

\bibitem{ancajas2014fort}
D.~M. Ancajas, K.~Chakraborty, and S.~Roy, ``Fort-{N}o{C}s: {M}itigating the threat of a compromised {N}o{C},'' in \emph{Proceedings of the 51st Annual Design Automation Conference}, 2014, pp. 1--6.

\bibitem{xilinx_idf}
S.~Pitaka, ``Isolation {D}esign {F}low for {X}ilinx 7 {S}eries {FPGA}s or {Z}ynq-7000 {S}o{C}s ({V}ivado {T}ools),'' AMD, Tech. Rep., 2020, https://docs.amd.com/v/u/en-US/xapp1222-idf-for-7s-or-zynq-vivado.

\bibitem{gouveia2022behind}
I.~P. Gouveia, M.~V{\"o}lp, and P.~Esteves-Verissimo, ``Behind the last line of defense: Surviving soc faults and intrusions,'' \emph{Computers \& Security}, vol. 123, p. 102920, 2022.

\bibitem{schmid2010decentralised}
U.~Schmid and A.~Steininger, ``Decentralised fault-tolerant clock pulse generation in {VLSI} chips,'' Sep.~7 2010, uS Patent 7,791,394.

\bibitem{sepulveda2017towards}
J.~Sep{\'u}lveda, A.~Zankl, D.~Fl{\'o}rez, and G.~Sigl, ``Towards protected {MPS}o{C} communication for information protection against a malicious {N}o{C},'' \emph{Procedia computer science}, vol. 108, pp. 1103--1112, 2017.

\bibitem{srivastava2005performance}
N.~Srivastava and K.~Banerjee, ``Performance analysis of carbon nanotube interconnects for {VLSI} applications,'' in \emph{ICCAD-2005. IEEE/ACM International Conference on Computer-Aided Design}.\hskip 1em plus 0.5em minus 0.4em\relax IEEE, 2005, pp. 383--390.

\bibitem{PBFT1999}
M.~Castro and B.~Liskov, ``Practical byzantine fault tolerance,'' in \emph{Proceedings of the Third Symposium on Operating Systems Design and Implementation}, ser. OSDI '99.\hskip 1em plus 0.5em minus 0.4em\relax USA: USENIX Association, 1999, p. 173–186.

\bibitem{zyzzyva2010}
\BIBentryALTinterwordspacing
R.~Kotla, L.~Alvisi, M.~Dahlin, A.~Clement, and E.~Wong, ``Zyzzyva: {S}peculative {B}yzantine {F}ault {T}olerance,'' \emph{ACM Trans. Comput. Syst.}, vol.~27, no.~4, jan 2010. [Online]. Available: \url{https://doi.org/10.1145/1658357.1658358}
\BIBentrySTDinterwordspacing

\bibitem{verissimo2003uncertainty}
P.~Ver{\'\i}ssimo, ``Uncertainty and predictability: {C}an they be reconciled?'' in \emph{Future Directions in Distributed Computing: Research and Position Papers}.\hskip 1em plus 0.5em minus 0.4em\relax Springer, 2003, pp. 108--113.

\bibitem{yeh1996triple}
Y.~C. Yeh, ``Triple-triple redundant 777 primary flight computer,'' in \emph{1996 IEEE Aerospace Applications Conference. Proceedings}, vol.~1.\hskip 1em plus 0.5em minus 0.4em\relax IEEE, 1996, pp. 293--307.

\bibitem{traverse2004airbus}
P.~Traverse, I.~Lacaze, and J.~Souyris, ``Airbus fly-by-wire: {A} total approach to dependability,'' in \emph{Building the Information Society}.\hskip 1em plus 0.5em minus 0.4em\relax Springer, 2004, pp. 191--212.

\bibitem{mancini1986modular}
L.~Mancini, ``Modular redundancy in a message passing system,'' \emph{IEEE Transactions on Software Engineering}, no.~1, pp. 79--86, 1986.

\bibitem{kopetz2003time}
H.~Kopetz and G.~Bauer, ``The time-triggered architecture,'' \emph{Proceedings of the IEEE}, vol.~91, no.~1, pp. 112--126, 2003.

\bibitem{Verissimo2006}
\BIBentryALTinterwordspacing
P.~E. Ver\'{\i}ssimo, ``Travelling through {W}ormholes: {A} {N}ew {L}ook at {D}istributed {S}ystems {M}odels,'' \emph{SIGACT News}, vol.~37, no.~1, p. 66–81, mar 2006. [Online]. Available: \url{https://doi.org/10.1145/1122480.1122497}
\BIBentrySTDinterwordspacing

\bibitem{kapitza2012cheapbft}
R.~Kapitza, J.~Behl, C.~Cachin, T.~Distler, S.~Kuhnle, S.~V. Mohammadi, W.~Schr{\"o}der-Preikschat, and K.~Stengel, ``Cheap{BFT}: {R}esource-efficient {B}yzantine fault tolerance,'' in \emph{Proceedings of the 7th ACM european conference on Computer Systems}, 2012, pp. 295--308.

\bibitem{pagonis2023increasing}
G.~Pagonis, V.~Leon, D.~Soudris, and G.~Lentaris, ``Increasing the {F}ault {T}olerance of {COTS} {FPGA}s in {S}pace: {SEU} {M}itigation {T}echniques on {MPS}o{C},'' in \emph{International Symposium on Applied Reconfigurable Computing}.\hskip 1em plus 0.5em minus 0.4em\relax Springer, 2023, pp. 215--229.

\bibitem{tabish2021next}
R.~Tabish, ``Next-generation safety-critical systems using {COTS} based homogeneous multi-core processors and heterogeneous {MPS}o{C}s,'' Ph.D. dissertation, University of Illinois at Urbana-Champaign, 2021.

\bibitem{alcaide2022unboxing}
S.~Alcaide, G.~Cabo, F.~Bas, P.~Benedicte, F.~Mazzocchetti, F.~Cazorla, and J.~Abella, ``Unboxing the {S}and: on {D}eploying {S}afety {M}easures in the {P}rogrammable {L}ogic of {COTS} {MPS}o{C}s,'' in \emph{11th European Congress Embedded Real Time Systems (ERTS 2022)}, 2022.

\bibitem{nicholas2021secure}
G.~S. Nicholas, A.~S. Siddiqui, S.~R. Joseph, G.~Williams, and F.~Saqib, ``A secure boot framework with multi-security features and logic-locking applications for reconfigurable logic,'' \emph{Journal of Hardware and Systems Security}, vol.~5, no.~3, pp. 260--268, 2021.

\bibitem{Turan2021}
\BIBentryALTinterwordspacing
F.~Turan and I.~Verbauwhede, ``Trust in {FPGA}-{A}ccelerated {C}loud {C}omputing,'' \emph{ACM Comput. Surv.}, vol.~53, no.~6, dec 2020. [Online]. Available: \url{https://doi.org/10.1145/3419100}
\BIBentrySTDinterwordspacing

\bibitem{guneysu2012securely}
T.~G{\"u}neysu, I.~Markov, and A.~Weimerskirch, ``Securely sealing multi-{FPGA} systems,'' in \emph{International Symposium on Applied Reconfigurable Computing}.\hskip 1em plus 0.5em minus 0.4em\relax Springer, 2012, pp. 276--289.

\bibitem{owen2018autonomous}
D.~Owen~Jr, D.~Heeger, C.~Chan, W.~Che, F.~Saqib, M.~Areno, and J.~Plusquellic, ``An autonomous, self-authenticating, and self-contained secure boot process for field-programmable gate arrays,'' \emph{Cryptography}, vol.~2, no.~3, p.~15, 2018.

\end{thebibliography}

\appendix
\label{sec:proofs}

\subsection{Safety}
In the context of our protocol, proving that Samsara is safe refers to ensuring Integrity is preserved.

\paragraph{Bootstrapping:}
The Controller, being a simple and easily-verifiable trusted-trustworthy ASIC, cannot be tampered with and follows only the designed logic at all times. During the boot phase, it signals the MP-Boot to start execution of the Bootloader code. This microprocessor is the only software processing unit capable of accessing the PL and its memory and is triggered only by the Controller (\textit{Bootstrapping phase, step 2}). Since no other software is allowed to run on MP-Boot, the configuring API is only called by the Bootloader and Tileloader, which are kept in Tamper-Resistant Storage in encrypted form and authenticated by the Controller. Therefore, the initial configuration of the PL is only executed at boot time as demanded by the Controller and not triggered arbitrarily by malicious code. The same applies to the Tileloader during Rejuvenation. (\textit{Bootstrapping phase, step 3})

Full and partial bitstreams are stored in encrypted form in the Softcore Library and authenticated when loaded into the PL by modules such as the AES-GCM engine, present in, e.g, Xilinx UltraScale+ devices as part of the configuration Security Unit (CSU). Without knowledge of the AES-GCM key, a bitstream cannot be modified or forged. Given authenticated bitstreams can still be malicious in nature or suffer faults when deployed into the PL, the necessity for Rejuvenation remains.

Furthermore, techniques such as those presented in~\cite{guneysu2012securely,owen2018autonomous} utilize the Xilinx Internal Configuration Access Port (ICAP) to read FPGA configuration memory at runtime and generate a hash for comparison against an expected hash with the goal of detecting runtime tampering, which the Bootloader and Tileloader can used for further verification.

Since the Controller, MP-Boot and Softcore Lib are tamperproof and the MPSoC designed to grant only access to the required components, it follows that the Compute Platform and tiles work as defined.

\paragraph{Execution:}
During PL-side executions, to ensure message requests are not dropped, the Controller assigns an \emph{unique ID} to each message, which is incremented by a hardware logic counter it implements. This counter is monotonic and used to associate sequence numbers to each operation (\textit{Execution phase, step 2}), similarly to USIG in MinBFT~\cite{MinBFT2013} . Given only the Controller has write access to its specific set of BRAM memory (PLM-C) from which the Tiles will read requests from, it is guaranteed that requests are originated from the Controller. Additionally, requests are hashed for integrity and verified by the Tiles, which, upon execution of the request, place a reply with the same ID in hashed form in their own BRAM memory to which only they have write access, proving its origin. Only replies with the corresponding ID in the corresponding address offset (dictated by the ID) are accepted by the Controller. The fact that messages are hashed means they cannot be forged by the bus or a malicious PL IP.

Finally, the Controller matches replies with the same \emph{unique ID} from all tiles, forwarding the result to the invoking application only in case of majority ($2f + 1$ or $f + 1$). Otherwise, Rejuvenation is triggered based on the chosen policy. Given this processing is done by hardware logic alone, the Controller is trusted to provide the correct result to the application (\textit{Execution phase, step 4}).

\paragraph{Rejuvenation:}
Rejuvenation works similarly to Bootstrapping, as the Tileloader is triggered by the Controller upon mismatch and uses a similar API to load the partial or full bitstreams into the PL. The Controller uses the parameters defined in Config that dictate the type of softcore/IP and version to load, which is stored in the TRS (\textit{Rejuvenation phase, step 2}). Even in the eventuality that a wrong bitstream version was loaded, the execution phase would trigger a mismatch in the Controller, leading to another Rejuvenation. Softcore/IP digests are also registered in the Controller TRS to ensure no foreign bitstream is loaded.

Finally, state transfer is guaranteed to be correct since the Controller stores the hashed state during the Execution phase, after successfully matching each request reply, in the PLM-C BRAMs to which only it has write access. The state retains all the previous matching replies delivered, while non-delivered requests should be replayed by the application. After a successful Rejuvenation, Tiles read the hashed state from the Controller's PLM-C into their own BRAM so that they can keep building upon it if required by a stateful application. No other Tile can forge state in another Tile's BRAM given no read or write access is provided. Furthermore, the Controller issues new Requests only when Tiles are loaded and state transfer ends, as signaled by the Tileloader and the Tiles themselves. In the event of a full PL Rejuvenation, which wipes the PLM-C BRAMs, the Controller first takes a Checkpoint, saves it to the on-chip SRAM and only then triggers the Tileloader.

\subsection{Liveness}
We detail now the informal proof that availability and termination are achieved.

\paragraph{Bootstrapping:}
The Bootloader and Tileloader are executed as bare-metal applications. No other application is allowed to execute in the MP-Boot, therefore they will never be preempted. The Controller signals when execution shall take place via an interrupt and MP-Boot's interrupt input port is connected only to the Controller. Similarly, the Config TRS is connected only to the Controller, which is trusted not to flood it, and configurations are sent to the MP-Boot by an exclusive channel connecting only the Controller and the MP-Boot (\textit{Bootstrapping phase, step 3}). Since no other element of the MPSoC has access to these configurations, they shall always be available. The Controller only triggers the Bootloader at boot time, before the PL starts receiving requests; and only signals the MP-Boot to execute the Tileloader after detecting a mismatch in replies. A mismatch is observed after the PL tiles have executed and replied, as such, before sending new requests, the Controller triggers the Tileloader, which reconfigures the PL (partially or fully) while Tiles are waiting for a new request (\textit{Bootstrapping phase, step 6}). Reconfiguration is preemptive and will not fail even if the Tile is executing some rogue code or finishing logging the request in its local memory (in stateful applications). A corrupted log is also irrelevant as it is never sent to any other Tile or the Controller. The latter records its own log, which is used for state transfer and is always available in the PLM-C or the on-chip SRAM.

In order to prevent Rejuvenation from stalling, the Controller sets a timer and waits for Ready responses from the corresponding rejuvenated tiles and the Bootloader, Tileloader and Tiles notify the Controller as being ready when Rejuvenation has finished. This ensures normal operation is resumed as soon as Rejuvenation is done and that after a pre-defined time limit, if a Tile is not ready, Rejuvenation is triggered again and the Tile is swapped for a diverse one.

\paragraph{Execution:}
The Controller uses parallelized logic to receive requests and place them in queue while it processes replies from the Tiles. Nevertheless, to ensure correct operation, the Controller only starts accepting requests when its status is Ready, meaning requests are not accepted while the PL is Rejuvenating. Jamming the Controller may be attempted by faulty applications, however, due to hardware parallelism, the input ports and request verification logic are independent of the logic used to handle requests (i.e., sending requests to replicas, waiting for their reply, verifying agreement, etc) and trigger Rejuvenation. The Controller can ignore requests sent by an application if they are sent with a time interval less than a determined delta or above a maximum threshold number within a time window.

Additionally, requests, when issued by the Controller, are readily available to the Tiles, due to being written in the PLM-C, to which Tiles have read access. Each Tile has its own read-only access channel to the PLM-C memory controller, granting them exclusive use of the bus and ensuring there is no contention. The same goes for Tiles writing replies in their own local BRAMs and the Controller reading them. No replays happen since each message has its own \emph{unique ID} and is placed in a specific address offset according to the said \emph{unique ID} (\textit{Execution phase, step 3}). The Controller sets a timer to wait for responses, as highlighted in step 2 of Section~\ref{sec:exec_phase}, to make sure faulty replicas to not delay agreement indefinitely.

The Controller considers an operation successful only if it finds majority matches, which happens since a maximum of $f$ faulty Tiles are assumed. Otherwise Rejuvenation will happen if one of the Tile is faulty and another is slow). The Controller always delivers the matched reply to the waiting application (\textit{Execution phase, step 4}).

\paragraph{Rejuvenation:}
This is only invoked by the Controller that is trusted either in Reactive or Proactive mode. Thus, the system cannot keep rejuvenating except when the Controller notifies the MP-Boot, when faults happen or on well-defined periods (in case of proactive-mode) (\textit{Rejuvenation phase, step 2}). In the former, \proto invokes the Tileloader using the given Config (that is always available and secure) and parameters (full-mode or partial-mode). The same is observed in Bootstrapping with the exception of state-transfer. This is available since the Controller stores the state in a dedicated BRAM to which Tiles only have read access through dedicated channels. Furthermore, state is hashed so that it is copied intact to the Tile's BRAM. Finally, Tiles signal they are ready when state transfer is done before the Controller's timer is out, in which case it Rejuvenates. After Rejuvenation completes, the Controller's status is set to ready and the Compute Platform is ready to execute new requests (\textit{Rejuvenation phase, step 5}).

\end{document}